\documentclass[10pt,journal,compsoc]{IEEEtran}

\ifCLASSOPTIONcompsoc
  \usepackage[nocompress]{cite}
\else
  \usepackage{cite}
\fi

\ifCLASSINFOpdf
  \usepackage[pdftex]{graphicx}
\else
\fi

\usepackage{amsmath}

\usepackage{algorithm}

\usepackage{algpseudocode}

\usepackage{array}

\ifCLASSOPTIONcompsoc
 \usepackage[caption=false,font=footnotesize,labelfont=sf,textfont=sf]{subfig}
\else
 \usepackage[caption=false,font=footnotesize]{subfig}
\fi
\usepackage{multirow}
\usepackage{booktabs}

\usepackage{url}

\newcommand{\squishlist}{
  \begin{list}{$\bullet$}
    { \setlength{\itemsep}{0pt}      \setlength{\parsep}{0pt}
      \setlength{\topsep}{0pt}       \setlength{\partopsep}{0pt}
      \setlength{\leftmargin}{1em} \setlength{\labelwidth}{1em}
      \setlength{\labelsep}{0.5em} } }
\newcommand{\squishlisttwo}{
  \begin{list}{$\bullet$}
    { \setlength{\itemsep}{0pt}    \setlength{\parsep}{0pt}
      \setlength{\topsep}{0pt}     \setlength{\partopsep}{0pt}
      \setlength{\leftmargin}{2em} \setlength{\labelwidth}{1.5em}
      \setlength{\labelsep}{0.5em} } }
\newcommand{\squishlistend}{
    \end{list}  }

\begin{document}
\title{Semantic-Aware Fuzzing: An Empirical Framework for LLM-Guided, Reasoning-Driven Input Mutation}

\author{Mengdi~Lu%
        , Steven~Ding, %
        Furkan~Alaca, %
        and~Philippe~Charland
\IEEEcompsocitemizethanks{
\IEEEcompsocthanksitem M. Lu and F. Alaca are with the School of Computing, Queen's University, Kingston, ON, Canada. 
\IEEEcompsocthanksitem S. Ding is with McGill University, Montreal, QC, Canada. 
\IEEEcompsocthanksitem P.Charland is with Mission Critical Cyber Security Section, Defence R\&D Canada. \protect\\} 

\thanks{Manuscript received September, 2025}}

\IEEEtitleabstractindextext{%
\begin{abstract}

Security vulnerabilities in Internet-of-Things (IoT) devices, mobile platforms, and autonomous systems remain critical. Traditional mutation-based fuzzers—while effectively explore code paths—primarily perform byte- or bit-level edits without semantic reasoning. Coverage-guided tools such as AFL++ rely on dictionaries, grammars, and splicing heuristics to impose shallow structural constraints, leaving deeper protocol logic, inter-field dependencies, and domain-specific semantics unaddressed. Conversely, reasoning-capable large language models (LLMs) have potentials to leverage human knowledge embedded during pretraining to understand input formats, respect complex constraints, and propose targeted mutations, much like an experienced reverse engineer or testing expert. However, without ground truth for “correct” reasoning in mutation generation, supervised fine-tuning is impractical, motivating explorations of off-the-shelf LLMs using prompt-based few-shot learning.
To bridge this gap, we present an open-source microservices framework that integrates reasoning LLMs with AFL++ on Google’s FuzzBench, addressing the asynchronous execution and divergent hardware demands (GPU- vs. CPU-intensive) of LLMs and fuzzers. We evaluate four research questions: (R1) How can reasoning-based LLMs be integrated into the fuzzing mutation loop? (R2) Do few-shot prompts yield higher-quality mutations than zero-shot? (R3) Can off-the-shelf reasoning models improve fuzzing directly via prompt engineering? and (R4) Which open-source reasoning LLMs perform best under prompt-only conditions? Experiments with Llama3.3, Deepseek-r1-Distill-Llama-70B, QwQ-32B, and Gemma3 highlight Deepseek-r1-Distill-Llama-70B as the most promising. Mutation effectiveness depends on prompt complexity and model choice rather than shot count alone. Response latency and throughput bottlenecks remain key obstacles. Our framework, released as open-source, supports reproducibility and community extension. Future directions include dynamic scheduling, lightweight feedback, and scalable deployment.

\end{abstract}

}

\maketitle

\IEEEdisplaynontitleabstractindextext

\IEEEpeerreviewmaketitle

\IEEEraisesectionheading{\section{Introduction}\label{sec:introduction}}

\IEEEPARstart{E}{ach} year, tens of thousands of new Common Vulnerabilities and Exposures (CVEs) are catalogued in the National Vulnerability Database (NVD), highlighting a rapidly expanding attack surface across Internet-of-Things (IoT) devices, mobile platforms, and autonomous systems \cite{mell2002cve,nvd_nist,SASI2024455,xu2014iot}. Manual code review—including labor-intensive reverse engineering of binaries—struggles to keep pace with this growth, as expert analysts can only examine a limited fraction of complex firmware, applications, or embedded controllers within reasonable timeframes. In contrast, fuzzing—an automated, “shift-right” testing approach that requires no source code and instead exercises compiled binaries with malformed or randomized inputs—has become one of the most effective vulnerability discovery techniques, responsible for identifying a majority of high-severity bugs in large software projects \cite{sutton2007what,most-eff1,most-eff2,most-eff3,most-eff4,10580893}. Coverage-guided, mutation-based fuzzers such as AFL++ \cite{10.5555/3488877.3488887} combine lightweight instrumentation with seed-based mutations to rapidly explore execution paths, and have demonstrated success against IoT firmware \cite{afliot,touqir2025fuzzing}, mobile applications \cite{mobilefuzz}, and autonomous driving stacks \cite{kim2022drivefuzz}. Despite these advances, existing fuzzers still rely on largely blind or heuristic-driven mutations that struggle to penetrate deep protocol logic and intricate input formats, which motivates further research into semantically informed mutation strategies.

Recent LLM-based fuzzing efforts have made impressive progress in applying prompt engineering to generate both initial seeds and targeted mutations for structured inputs. Fuzz4All demonstrated this approach on programming-language grammars \cite{10.1145/3597503.3639121}, PromptFuzz extended it to library APIs \cite{10.1145/3658644.3670396}, and CHATAFL showcased interactive mutation refinement through chat-style prompts when fuzzing protocols \cite{meng2024llmfuzz}. These works showcases the promise of LLMs to automate complex input synthesis. However, these approaches treat the model as a black box, focusing solely on input-to-output mapping and omitting the intermediate reasoning steps that ground high-quality generation. Chain-of-thought (COT) reasoning has been shown to improve LLM fidelity, reduce hallucinations, and enhance output diversity by making the model’s analytical process explicit.

Meanwhile, systems such as LLAMAFUZZ \cite{zhang2024llamafuzz} employ supervised fine-tuning on AFL++. LLAMAFUZZ derives “good” mutations, which inherently restricts the model’s creativity at existing mutation heuristics and requires costly labeled data. In contrast, our work explores whether prompting off-the-shelf reasoning LLMs—without any additional fine-tuning—can leverage their latent human-knowledge representations to generate semantically rich, novel mutations beyond AFL++’s conventional strategies.

Building on these observations, we hypothesize that reasoning-enabled LLMs can approximate the analytical workflow of an expert reverse engineer by examining an input’s structure, inferring inter-field dependencies, and applying domain knowledge to craft targeted mutations, rather than merely relying on surface‐level edits. By exposing the model’s internal “chain-of-thought” \cite{cot} through empirically designed prompts, we aim to unlock this latent reasoning capability, reduce blind spots in protocol logic, and minimize redundant or invalid mutations. Since there is no definitive “ground truth” for how such a reasoning process should proceed, and because supervised fine-tuning on AFL++ outputs would inherently limit the model to existing heuristics, we focus our empirical study on zero-shot and few-shot prompting—where the number of in-context examples provided to the model defines the “shot” count. This setup allows us to assess the raw reasoning power and creative mutation strategies of off-the-shelf LLMs without capping their potential or incurring the high cost of labeled data. We structure our empirical investigation around four research questions:

\squishlist
  \item \textbf{Research Question R1}: How can reasoning-based LLMs be integrated into the mutation loop of a coverage-guided fuzzer? This requires reconciling the asynchronous execution pace and distinct hardware requirements of CPU-centric fuzzers and GPU-backed LLMs without impairing overall throughput.
  \item \textbf{Research Question R2}: Does providing few-shot examples in prompts lead to higher-quality, more semantically informed mutations than zero-shot prompts? This evaluates whether example-driven prompting consistently enhances mutation validity and diversity over minimal prompt designs.
  \item \textbf{Research Question R3}: Can off-the-shelf reasoning LLMs improve fuzzing effectiveness through prompt-based reasoning alone? This explores whether an approach using only prompts, without any fine-tuning, can generate semantically meaningful mutations that achieve higher coverage or uncover more bugs.
  \item \textbf{Research Question R4}: Which open-source LLM yields the best performance guided solely by prompt engineering? This compares models to identify whose latent knowledge and reasoning capabilities translate most effectively into valid, high-quality test-case mutations.
\squishlistend

Building on these questions, we contribute an open-source, microservices-based framework that connects AFL++ to off-the-shelf reasoning LLMs via Redis and Docker, effectively harmonizing CPU- and GPU-driven components without fine-tuning. Our framework is packaged for automated deployment on Google’s FuzzBench, enabling reproducible, large-scale evaluation across diverse binary targets \cite{10.1145/3468264.3473932}. To the best of our knowledge, we present the first systematic study of applying reasoning-capable LLMs to mutation-based binary fuzzing using prompt engineering alone, comparing zero-, one-, and three-shot strategies to quantify their effects on mutation validity, code coverage, and crash discovery \cite{KNOTH2024100225}. In tandem, we empirically benchmark four state-of-the-art open-source reasoning models—Llama 3.3, DeepSeek-r1-Distill-Llama-70B, QwQ-32B, and Gemma 3—assessing their out-of-the-box mutation performance in coverage-guided fuzzing \cite{2024arXiv240721783G, bau2024gradient,gemma3,qwen32b2024}.  Finally, we analyze practical limitations—model latency, throughput trade-offs, and semantic depth—and outline directions for dynamic scheduling, lightweight feedback loops, and scalable deployment in LLM-driven fuzzing.

\section{Methodologies}
We propose a solution that incorporates a LLM as an independent service to assist the mutation stage of a grey-box, code-coverage-based fuzzer. Integrating LLM with the fuzzer (\textbf{R1}) introduces two major challenges. While LLMs are capable of advanced reasoning, they typically experience high response latency \cite{llmspeed} to process input and generate responses, in contrast to the high execution throughput of modern fuzzers like AFL++ \cite{10.5555/3488877.3488887}, which can process over $3000$ executions per second \cite{aflspeed}. Embedding an LLM directly in the fuzzing loop would degrade the fuzzing speed, making it difficult to maintain fuzzing efficiency (\textbf{Challenge 1}).
To preserve the performance of the fuzzer, we decouple the fuzzing and LLM mutation processes into two distinct components: \emph{a C-based fuzzer component} and \emph{a Python-based LLM-guided component}. These components are packaged and deployed independently, creating a synchronization challenge (\textbf{Challenge 2}) for communication between these two components, due to language differences and distributed execution environments.
\begin{figure}[t]
    \centering
    \includegraphics[width=0.45\textwidth]{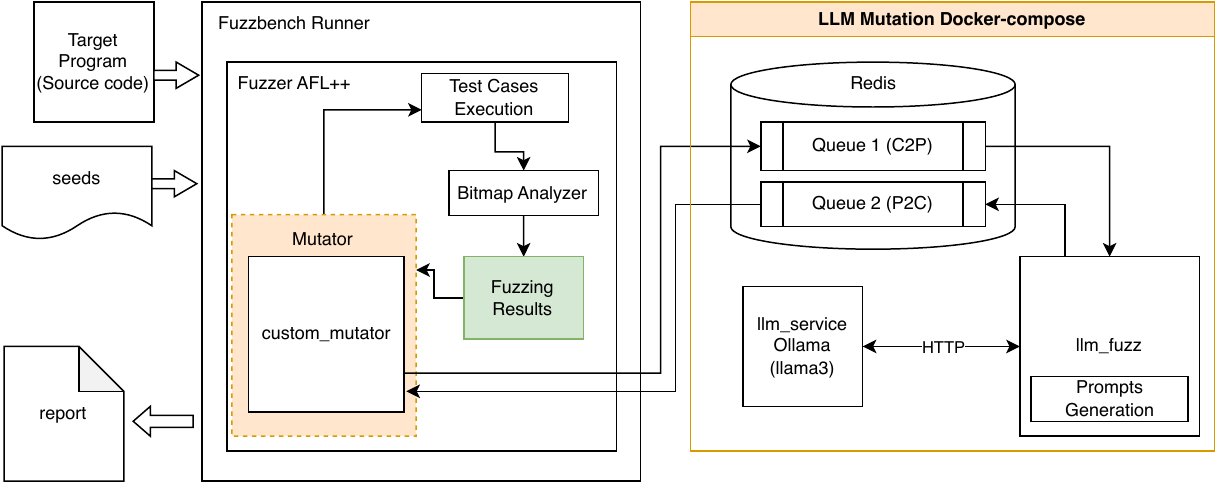}
    \caption{LLM guided fuzzing architecture overview}
    \label{Fig:Overview}
\end{figure}

\begin{figure}[t]
    \centering
    \includegraphics[width=0.45\textwidth]{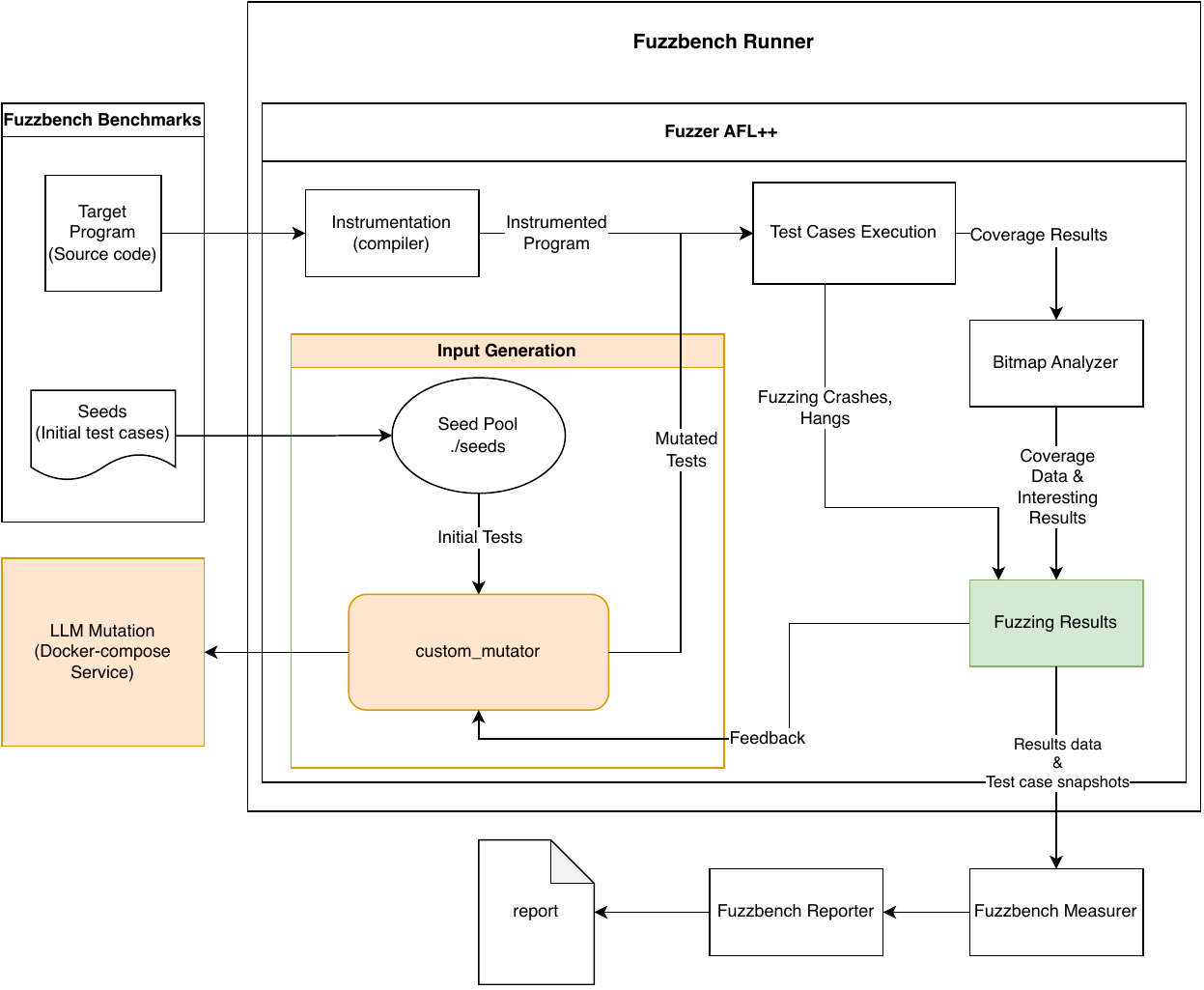}
    \caption{LLM guided fuzzing architecture: Fuzzer component - AFL++}
    \label{Fig:FuzzerComponent}
\end{figure}

\begin{figure}[t]
    \centering
    \includegraphics[width=0.45\textwidth]{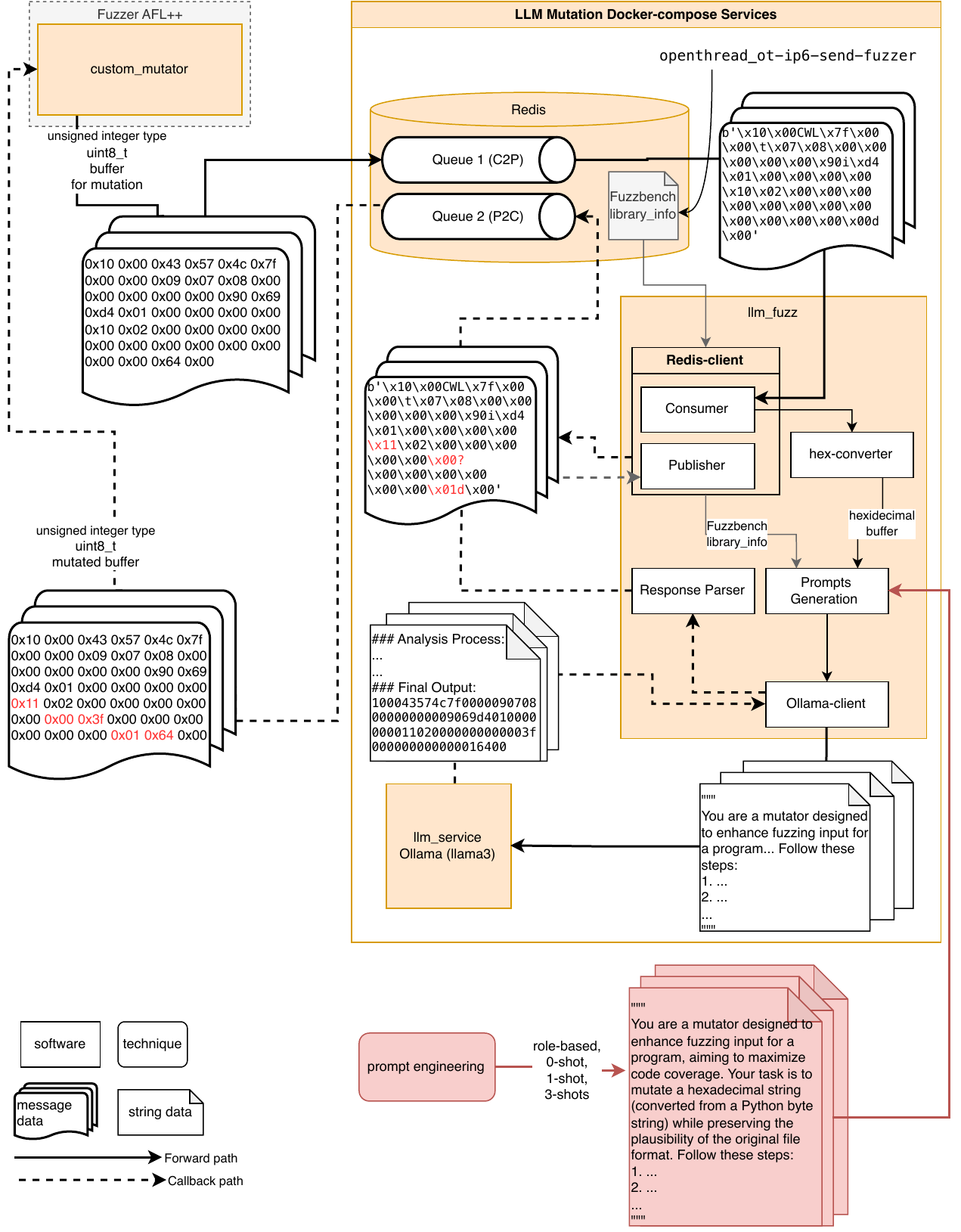}
    \caption{LLM guided fuzzing architecture: LLM Mutation component}
    \label{Fig:MutationComponent}
\end{figure}

To enable communications between the fuzzer and LLM mutating service, we extended AFL++'s mutation logic through a custom mutation hook, \texttt{custom\_mutator} interface (\textbf{Solution 2}). The implementation allows the fuzzer to invoke LLM-generated mutations in a selective and asynchronous manner.  The LLM mutation runs as a standalone Docker Compose service comprising three core components: a \emph{message broker (Redis)}, \emph{Ollama (llm\_service)}, and a \emph{LLM prompt generator (llm\_fuzz)}. To ensure efficient and reliable communications, multiple named message queues are established in the Redis server (\textbf{Solution 1}).
When LLM-generated mutated test cases are available in the Redis queue, the fuzzer consumes and uses them; otherwise, it defaults to built-in mutation logic. A high-level integration and architectural overview is shown in Figure \ref{Fig:Overview}, \ref{Fig:FuzzerComponent} and \ref{Fig:MutationComponent}.

\subsection{Infrastructure}
The proposed system is built on \emph{FuzzBench} \cite{10.1145/3468264.3473932}—an open-source fuzzing evaluation platform developed by Google—providing a standardized environment for evaluating fuzzers through automated benchmark execution, result analysis, and report generation. 
Our system, consisting of a custom fuzzer and an LLM mutation module, leverages this infrastructure to evaluate LLM-guided fuzzing across multiple benchmarks. The following subsection outlines the integration process and associated implementation challenges.

\subsubsection{Integrating Fuzzer into Fuzzbench}
While FuzzBench \cite{metzman2021fuzzbench} aims to simplify the evaluation and comparison of fuzzing techniques, the underlying architecture of building, deploying, executing fuzzers is inherently complex. As a result, integrating a custom fuzzer into the platform remains complex (\textbf{Challenge 3}). To incorporate our LLM-guided fuzzer, we begin by duplicating the packaging and deployment strategies from \emph{AFL++} in Fuzzbench, extending them for our implementation. The new fuzzer is added to the \texttt{./fuzzers} directory, and Docker containers are built using Fuzzbench's base image for both the custom builder and runner. However, additional modifications—such as new deployment scripts, updated configurations for custom mutation strategies, and fixes for Python version incompatibility and HTML report generations—are required to support both the fuzzer and LLM components. Through these modifications (\textbf{Solution 3}), the system is successfully integrated into FuzzBench, enabling reproducible experiments across benchmarks. Each trial produces detailed coverage reports for a systematic evaluation of the fuzzer’s performance on the selected benchmarks.

\subsubsection{Fuzzer Component}
AFL++ \cite{10.5555/3488877.3488887}, known for its efficient instrumentation and mutation strategies, acts as the core fuzzer in the fuzzer component. As illustrated in Figure \ref{Fig:FuzzerComponent}, the \emph{target program} and the \emph{initial test cases}, so-called \emph{seed corpus}, are supplied by Fuzzbench. During the build stage, AFL++ instruments the target program during compilation, enabling real-time monitoring of code coverage throughout execution. These instrumented binaries are then used by the fuzzing engine to observe and collect coverage feedback. Finally, feedback from each execution, such as newly discovered code paths or crashes, is analyzed to guide future mutations and enhance the overall efficiency of the fuzzing process.  Major stages in our fuzzing process includes:
\begin{enumerate}
    \item \textbf{Input generation:} Initial seeds are stored in a seed pool and serve as the basis for generating test cases.
    \item \textbf{Mutation:} Test cases, which are input buffers, are mutated using a \emph{dual mutation strategy} implemented in \textit{custom\_mutator}, altering AFL++’s original methods. Our mutation includes (1) standard AFL++ mutation operators such as bit flipping, arithmetic operations, and dictionary substitutions \cite{10.5555/3488877.3488887}, alongside (2) an LLM-guided mutation service. The LLM service—running in Docker Compose with Redis, Ollama, and a prompt generation module—performs semantic-aware mutations by targeting input sections most likely to reveal new code paths based on their structure and content. Figure \ref{Fig:MutationComponent} illustrates the LLM mutation service architecture.

    \item \textbf{Test Case Execution:} Mutated test cases are executed against the instrumented target program, with AFL++ \cite{10.5555/3488877.3488887} generating coverage bitmaps to track paths exploration. Test cases that cause crashes or timeouts are stored in separate output queues, crashes and hangs, for further analysis.
    \item \textbf{Results analysis:} The coverage bitmap reflects the diversity and frequency of executed branch tuples. In the bitmap, test cases that explore new paths are considered as “interesting” and increase coverage; they are prioritized in future fuzzing rounds.
    \item \textbf{Feedback mechanism:} The fuzzer dynamically adjusts its mutation and scheduling strategy based on bitmap feedback. Interesting test cases are prioritized, and subsequent mutations are guided by how much coverage improvements they have.
    \item \textbf{Report generation:} When finishing fuzzing, FuzzBench \cite{10.1145/3468264.3473932} aggregates all outputs with its measurer module to calculate final code coverage, then sends the analyzed results to the reporter. The reporter module renders the information in HTML, and generates the report.
\end{enumerate}

\subsubsection{LLM Mutation Component}
The LLM-guided mutation component operates as a standalone service integrated into the fuzzing infrastructure. Its architecture and message processing workflow are illustrated in Figure \ref{Fig:MutationComponent}, using the benchmark \textit{openthread\_ot-ip6-send-fuzzer} as an example. This component is composed of three primary microservices:

\paragraph{Redis -- Message Broker and Context Store} Redis \cite{redis} is an open-source, in-memory data structure store commonly used as a message broker and cache system. It supports a key–value data model and belongs to the class of \textit{NoSQL} databases \cite{nosql}. 
Within this architecture, Redis bridges the AFL++ fuzzer (in \textit{C}), and the LLM mutation service (in \textit{Python}) through multiple uniquely named queues. The two primary queues are used: (1) \emph{C2P} (Client-to-Prompt), which stores messages containing input buffers published by AFL++ and consumed by the LLM mutation service; and (2) \emph{P2C} (Prompt-to-Client), which holds messages containing LLM-mutated buffers published by the LLM component and consumed by AFL++.
Moreover, Redis also maintains a persistent key-value pair identified as \textit{library\_info}, which stores metadata about the FuzzBench \cite{10.1145/3468264.3473932} benchmark libraries. This contextual information is essential for generating informed and context-aware prompts for the LLM during the mutation process. Thus, our message broker handles message queuing, manages shared states, enables asynchronous communications, and facilitates seamless integration between fuzzing and LLM components.

\paragraph{Ollama -- LLM Execution Engine} Ollama \cite{ollama} is an open-source platform designed for deploying and executing LLMs. It is containerized via Docker, which simplifies model deployment and isolation.
In this architecture, Ollama \cite{ollama} acts as the \emph{core inference engine} for the LLM-guided mutation process. It handles mutation requests, performs inference using the loaded model, and returns the generated outputs. While Ollama does not support direct fine-tuning, it leverages internal quantization and runtime optimizations aimed at reducing inference latency. These design choices make it a practical and efficient component for integrating into our fuzzing system.

\paragraph{\textit{llm\_fuzz} -- Prompt Generation and Message Management} The \textit{llm\_fuzz} service is a Python-based orchestration module responsible for message publishing and consuming, prompt generation and LLM response management. The service \textit{(1)} manages communication with \emph{Redis} for publishing and consuming messages, \textit{(2)} generates prompts and sends them to \emph{Ollama} for inference, and \textit{(3)} parses and handles responses returned by the LLM. 
Since LLMs are not inherently optimized for parsing complex binary formats such as object files (e.g. OTT), input buffers are first converted to \emph{hexadecimal representations} using a dedicated \emph{hex-converter}. This transformation ensures consistency in input format and allows the LLM to process the content more reliably.
Prompt generation within \textit{llm\_fuzz} is optimized using \emph{prompt design and engineering techniques}, which structure the input and context (e.g. \textit{library\_info}) to enhance the LLM’s mutation capabilities. The mutated output is found in the \emph{"Final Output"} section of the response received from Ollama. The section must be parsed by a custom \emph{response parser} to extract the actual mutated test case for the fuzzer. 
Overall, the \textit{llm\_fuzz} serves as a message and prompt manager, containing: a Redis client for data exchange, an Ollama client for LLM interaction, and a utility stack for hex conversion, prompt generation, and response parsing.

\subsection{Data Flow}
\begin{figure}[t]
    \centering
    \includegraphics[width=0.45\textwidth]{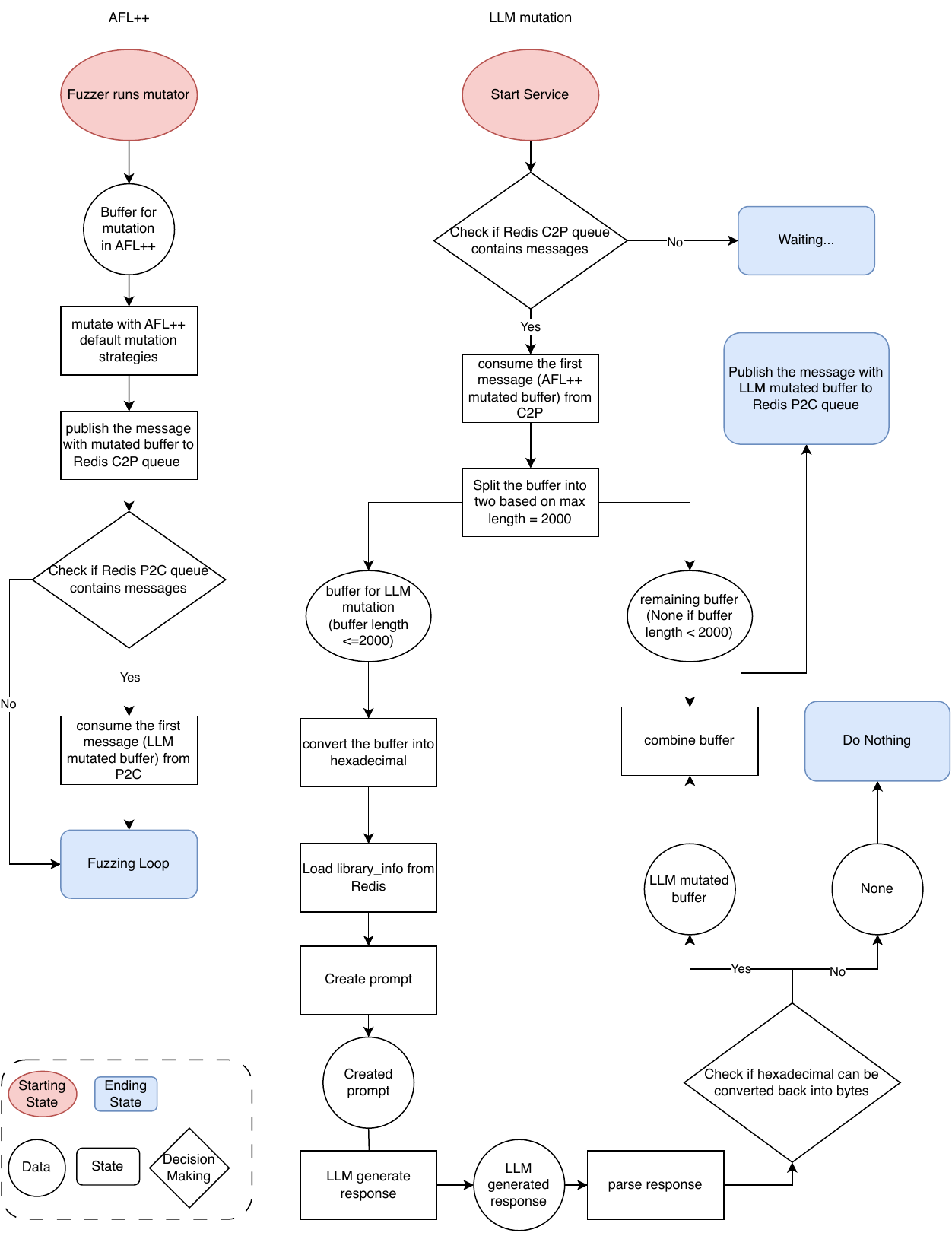}
    \caption{Data flow diagram of AFL++ component and LLM mutation component}
    \label{Fig:Dataflow}
\end{figure}

The proposed mutation method is outlined in \emph{Algorithm \ref{alg:llm_fuzzing}}. Complementing the pseudo-code, Figure \ref{Fig:Dataflow} presents a visualized data flow illustrating the interaction between AFL++ and LLM-guided mutation components in our proposed framework. The processing of input buffers during mutation is detailed in the following stages:
\squishlist
    \item \textbf{Message publishing in AFL++:} Input buffers, represented as \textit{uint8\_t} (\textit{unsigned 8-bit integers}), are initially mutated using AFL++'s default mutation strategies. The mutated buffers are then serialized and published to the Redis queue named \textit{C2P}.
    \item \textbf{Message consumption in \textit{llm\_fuzz}:}  The \textit{llm\_fuzz} service listens to the \textit{C2P} queue using the Redis client. When a message arrives, it is consumed as a \textit{Python byte strings}. Otherwise, the application enters a wait state.
    \item \textbf{Message processing in \textit{llm\_fuzz}:} Consumed messages in the \textit{llm\_fuzz} service are processed in two key steps:
    \squishlist
        \item \textbf{Buffer Splitting:} Buffers longer than \textit{2000} bytes are split to prevent memory overflow on CUDA-enabled systems. The first $2000$ bytes are mutated by the LLM, while the remaining segments are stored for recombination. Buffers shorter or equal to $2000$ bytes are used in full in the mutation, and the remaining segments are considered as \textit{empty}. This strategy prioritizes mutating the first segment as file format identifiers typically appear at the beginning of files.
        \item \textbf{Hexadecimal Conversion:} Buffers designated for mutation are converted to \textit{hexadecimal} strings to ensure compatibility with the LLM.
    \squishlistend
    
    \item \textbf{Prompt generation in \textit{llm\_fuzz}:} Refined prompts are created using prompt engineering, incorporating user contexts, including both the hexadecimal input buffer and the benchmark context information retrieved from Redis. These prompts are then sent to the Ollama service via an HTTP connection established by the Ollama client.
    \item \textbf{Response generation in LLM:} The LLM processes the prompt and returns a structured response containing two key sections: \textit{Analysis} and \textit{Final Output}. The \textit{Final Output} includes the mutated buffers in hex format, following strict formatting constraints.
    \item \textbf{Response parsing in \textit{llm\_fuzz}:} The mutated buffer is extracted and converted back from hexadecimal to Python byte string in the \textit{Final Output}. If a \textit{ValueError} occurs due to an odd-length hexadecimal string, a \textit{"0"} is appended to fix the format. Buffers that still fail are discarded and not passed to the fuzzer.
    \item \textbf{Message publishing in \textit{llm\_fuzz}:} Validly converted and recombined mutated buffers—including previously split segments—are published to the \textit{P2C} Redis queue, where they become available for AFL++ to consume and execute.
    \item \textbf{Message consumption in AFL++:} The AFL++ component continuously monitors the \textit{P2C} queue on the Redis server. When a message containing LLM-mutated buffers is received, AFL++ consumes the data, also in \textit{uint8\_t} format, and utilizes it in the fuzzing loop. Otherwise, AFL++ continues execution using its default mutations.
\squishlistend

\begin{algorithm}[t]
\caption{LLM-Guided Mutation-Based Fuzzing}
\label{alg:llm_fuzzing}
\emph{Input:} Seed corpus $\mathcal{C}_0$, Target program $\mathcal{F}$, Pre-trained LLM $\mathcal{M}_{\text{LLM}}$, Prompt shots $\mathcal{P}_k$ ($k \in \{0,1,3\}$), Fuzzing time interval $T$ \\
\emph{Output:} Set of interesting inputs $\mathcal{C}_{\text{mut}}$, Coverage results $\mathcal{R}_{\text{cov}}$, LLM response metrics $\mathcal{R}_{\text{log}}$
\begin{algorithmic}[1]
\State Initialize input queue $Q \gets \mathcal{C}_0$
\State Initialize coverage map $\mathcal{M}_{\text{cov}} \gets \emptyset$
\State Initialize $\mathcal{R}_{\text{log}}, \mathcal{C}_{\text{mut}}, \mathcal{Q}_{\text{crash}}, \mathcal{Q}_{\text{hang}} \gets \emptyset$
\For{$t = 1$ to $T$}
    \State Select input $x \sim Q$ using queue strategy 
    \State Split string $x$
    \Statex \hspace{1.35em} \(x_{length<=2000}, x_{length>2000} \gets \text{LengthSplitter}(x) \)
    \State Convert $x_{length<=2000}$ to hex string $h_x$
    \Statex \hspace{1.35em} \(h_x \gets \text{HexEncode}(x_{length<=2000})\)
    \State Format prompt $\pi \gets \text{FormatPrompt}(\mathcal{P}_k, h_x)$
    \State Query LLM $y \gets \mathcal{M}_{\text{LLM}}(\pi)$
    \If{$y = \emptyset$ or timeout}
        \State $x' \gets \text{Mutate}_{\text{AFL++}}(x)$
        \State \emph{continue}
    \EndIf
    \State Extract hex output $h_{x'} \gets \text{ParseLLM}(y)$
    \State Try decode $x' \gets \text{HexDecode}(h_{x'})$
    \If{decode fails}
        \State Log for FMR or HCER
        \State \emph{continue}
    \EndIf
    \State Re-combine strings $x' \gets Join(x', x_{length>2000})$
    \State Run $x'$ on $\mathcal{F}$: $(\mathcal{O}, \mathcal{C}) \gets \mathcal{F}(x')$
    \If{$\mathcal{O} = \text{CRASH}$}
        \State Add $x'$ to crash queue: $\mathcal{Q}_{\text{crash}} \gets \mathcal{Q}_{\text{crash}} \cup \{x'\}$
        \State \emph{continue}
    \ElsIf{$\mathcal{O} = \text{TIMEOUT}$}
        \State Add $x'$ to hang queue: $\mathcal{Q}_{\text{hang}} \gets \mathcal{Q}_{\text{hang}} \cup \{x'\}$
        \State \emph{continue}
    \ElsIf{$\text{isNewCoverage}(\mathcal{C})$}
        \State Add $x'$ to queue and corpus: 
        \Statex \hspace{2.85em} $Q \gets Q \cup \{x'\}$, $\mathcal{C}_{\text{mut}} \gets \mathcal{C}_{\text{mut}} \cup \{x'\}$
    \EndIf
    \State Log metrics $\mathcal{R}_{\text{log}} \gets \mathcal{R}_{\text{log}} \cup \{(x', \text{FMR}, \text{HCER}, \text{RDR})\}$
\EndFor
\State Generate final report $\mathcal{R}_{\text{cov}}$
\end{algorithmic}
\end{algorithm}
 
\subsection{Deployment}
The fuzzer component integrates with the FuzzBench framework \cite{10.1145/3468264.3473932} and is deployed alongside target benchmarks using Fuzzbench's standard pipeline. In contrast, the LLM mutation component is deployed independently using Docker Compose. The Docker Compose configuration defines and orchestrates \emph{Redis}, \emph{Ollama}, and \emph{\textit{llm\_fuzz}} microservices, which are initialized simultaneously using the \texttt{docker-compose up --build} command.
Once these services are active and the LLM models are loaded into Ollama, the fuzzing process can be initiated via our developed custom script, \texttt{run\_benchmark.sh}, which simplifies benchmark configuration and automates the building and deployment of the fuzzer component in Docker containers.
Using Docker Compose to deploy the LLM-guided mutation component provides three advantages:
\squishlist
    \item \textbf{Modularity:} The LLM-guided mutation service is fully decoupled from the fuzzer component, enabling independent deployment—even across different servers—and seamless LLM integration into fuzzing workflows.
    \item \textbf{Flexibility:} Both the fuzzer and LLM components can be easily configured via environment variables and command-line arguments in the deployment script, enabling quick adaptation to different benchmarks, fuzzing strategies, and model versions.
    \item \textbf{Sustainable Extension:} Our architecture allows other FuzzBench-integrated fuzzers to adopt the LLM service by using the \texttt{custom\_mutator} hook and configuring Redis queues appropriately.
\squishlistend

\subsection{Prompt Refinement} \label{prompt-eng}
In the LLM response generation stage, the output is expected to consist a clean hexadecimal string representing the mutated input—without any additional text, special characters, or explanatory content. Thus, a minimal prompt like: \textit{"Mutate the given  buffer \{ input\_buffer \} by using provided \{ library\_info \}, and generate only the mutated hexadecimal string."} can be sufficient.
However, using such simple prompts presents two significant limitations (\textbf{Challenge 4}): 
\begin{enumerate}
    \item \textbf{Inconsistent Output Format:} Due to the inherent stochastic nature of LLMs, outputs may deviate from the expected format, even with fixed parameters like temperature \cite{guo2022survey}. 
    \item \textbf{Limited Interpretability}: Simple prompts do not offer insights into the model's reasoning, making it difficult to analyze how results are derived.
\end{enumerate} 
To address these limitations and improve both output accuracy and interpretability, the prompts are refined using two core techniques: \emph{prompt design} and \emph{prompt engineering} (\textbf{Solution 4}).
Our prompt design strategies include:
\squishlist
    \item \textbf{Role playing:} Inspired by \cite{shanahan2023role}, the LLM is assigned a domain-specific role—a fuzzing expert—to improve task alignment and contextual relevance.
    \item  \textbf{Task clarification:} Prompts clearly and explicitly define the objective to instruct LLM about what to achieve. For example, my prompt states: \textit{“Your task is to mutate a hexadecimal string (converted from a Python byte string) while preserving the plausibility of the original file format.”} 
    \item  \textbf{Instruction specialization:} The mutation task is decomposed into fine-grained, step-by-step instructions, which includes: (1) \emph{analyze the input buffer}, (2) \emph{identify suitable mutation targets}, and (3) \emph{apply structured mutations}. Each step is further elaborated. For example, step (1) involves decoding the hexadecimal input, identifying structural components such as key fields, regions, and known library patterns, recognizing sections critical to format integrity, and identifying areas safe to mutation.
    \item \textbf{Section structuring:} Prompts are organized into logical segments: instruction, context, input data, and output format indicators. The response is structured into two parts: (1) \emph{"Analysis"}, a detailed rationale or “chain-of-thought” (COT) \cite{cot} explaining the mutation reasoning, which aids prompt refinement; (2) \emph{"Final Output"}, a strictly formatted hexadecimal string containing only the mutated buffer—free from explanations or extra characters—to ensure efficient parsing of LLM responses. 

\squishlistend 

Despite LLMs’ ability to understand complex prompts, their reliance on memory can lead to context loss in lengthy input sequences, resulting in less accurate or inconsistent responses. To mitigate this, a role-based message structuring strategy is used to semantically segment the prompt and clarify the intent of instructions. It is effective because LLMs can interpret input differently based on assigned roles. Accordingly, two distinct roles are incorporated into the \emph{prompt engineering strategy} in our research:
\begin{figure}[t]
    \centering
    \includegraphics[width=0.45\textwidth]{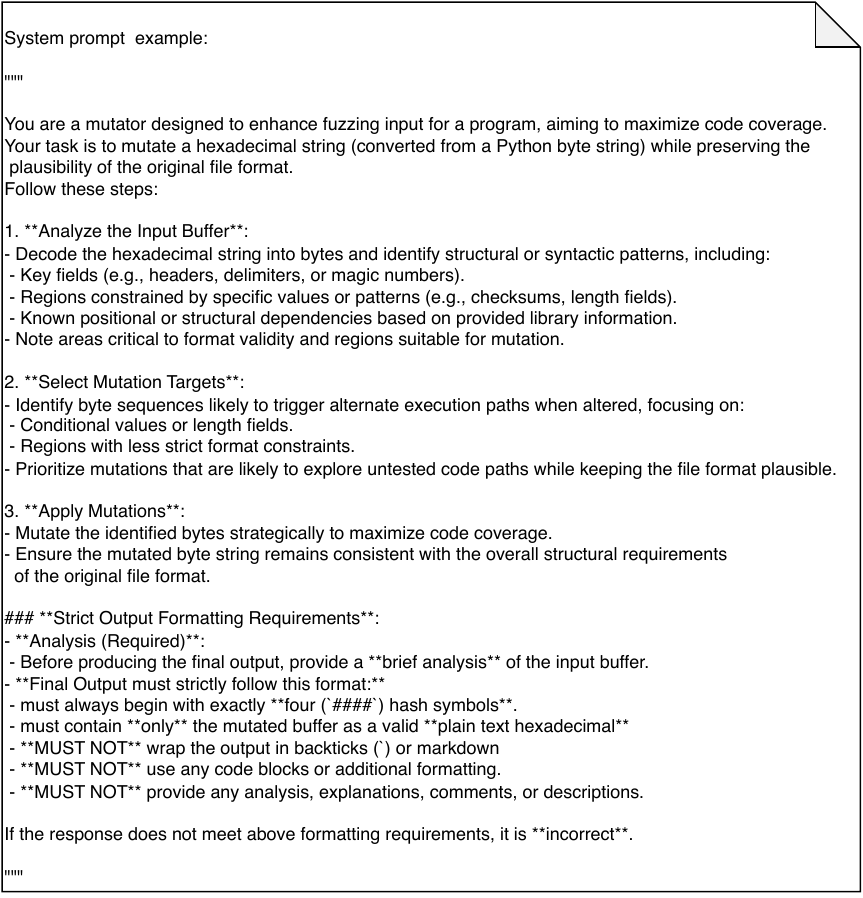}
    \caption{An example of a system prompt.}
    \label{fig:SystemPrompt}
\end{figure}
\subsubsection{System Role}
System-based prompts contain fundamental instructions to define the LLM's behavioral constraints and contextual scope. They establish limitations, rules, and ethical boundaries to guide response generation. In our study, system prompts consist of the assigned role, primary tasks, step-be-step instructions with contexts, and strict output format requirements. Since the fuzzer expects only the mutated buffer, a dedicated \emph{"Strict Output Formatting Requirements"} section is emphasized in the system prompt to enforce compliance. Figure \ref{fig:SystemPrompt} demonstrates a system-role-based prompt.

\subsubsection{User Role}
\begin{figure}[t]
    \centering
    \includegraphics[width=0.40\textwidth]{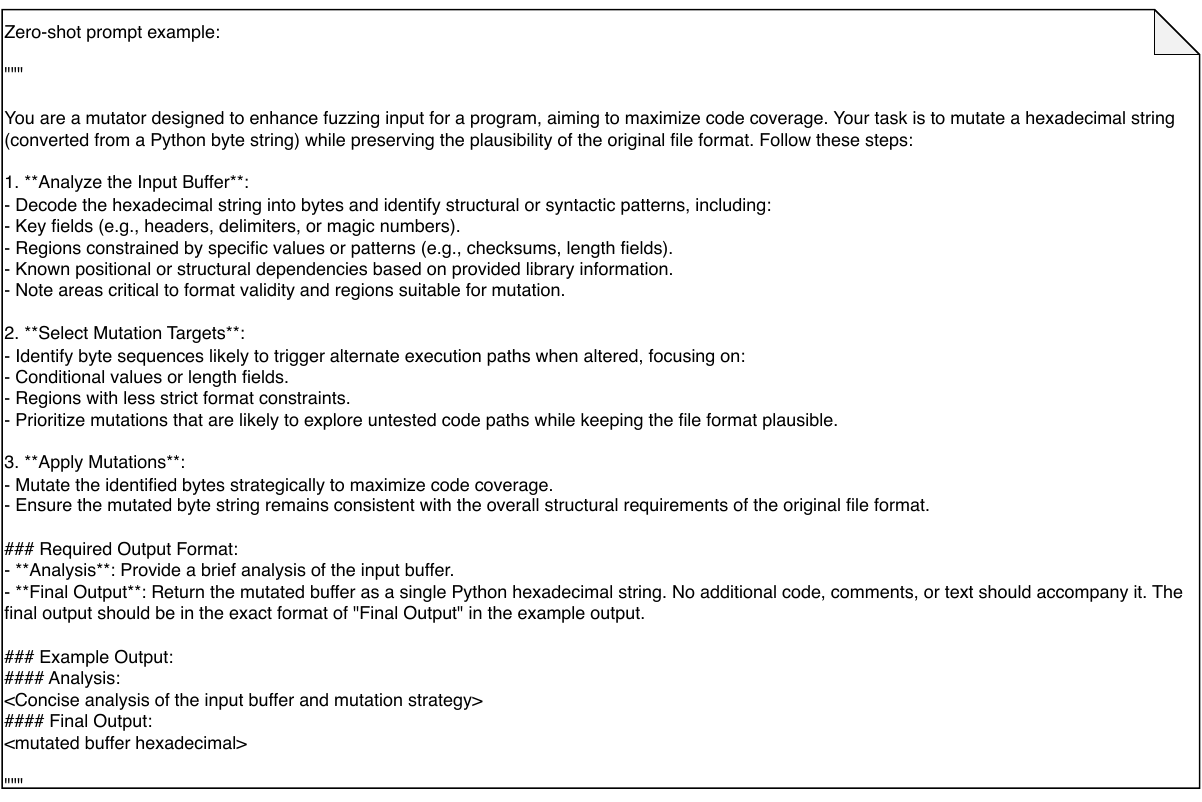}
    \caption{A structured prompt for zero-shot prompting.}
    \label{fig:ZeroShotPrompt}
\end{figure}

\begin{figure}[t]
    \centering
    \includegraphics[width=0.40\textwidth]{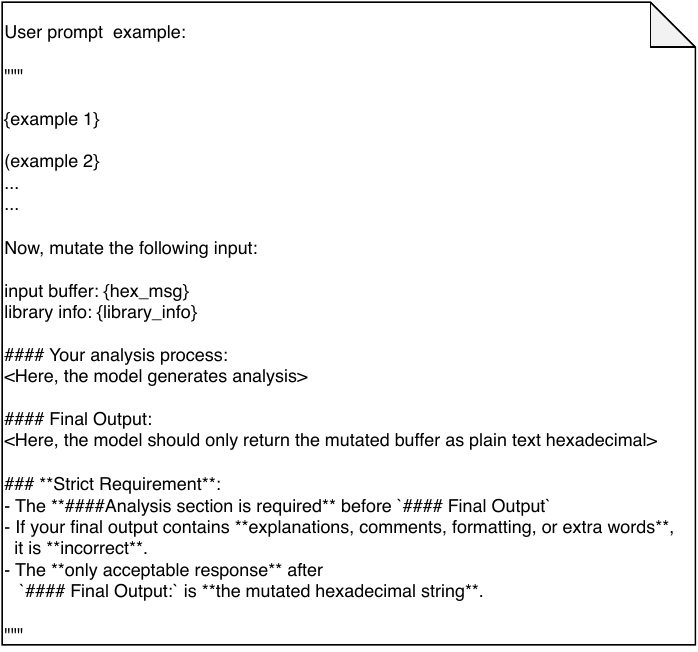}
    \caption{An example of a user prompt.}
    \label{fig:UserPrompt}
\end{figure}

\begin{figure}[t]
        \centering
        \includegraphics[width=0.48\textwidth]{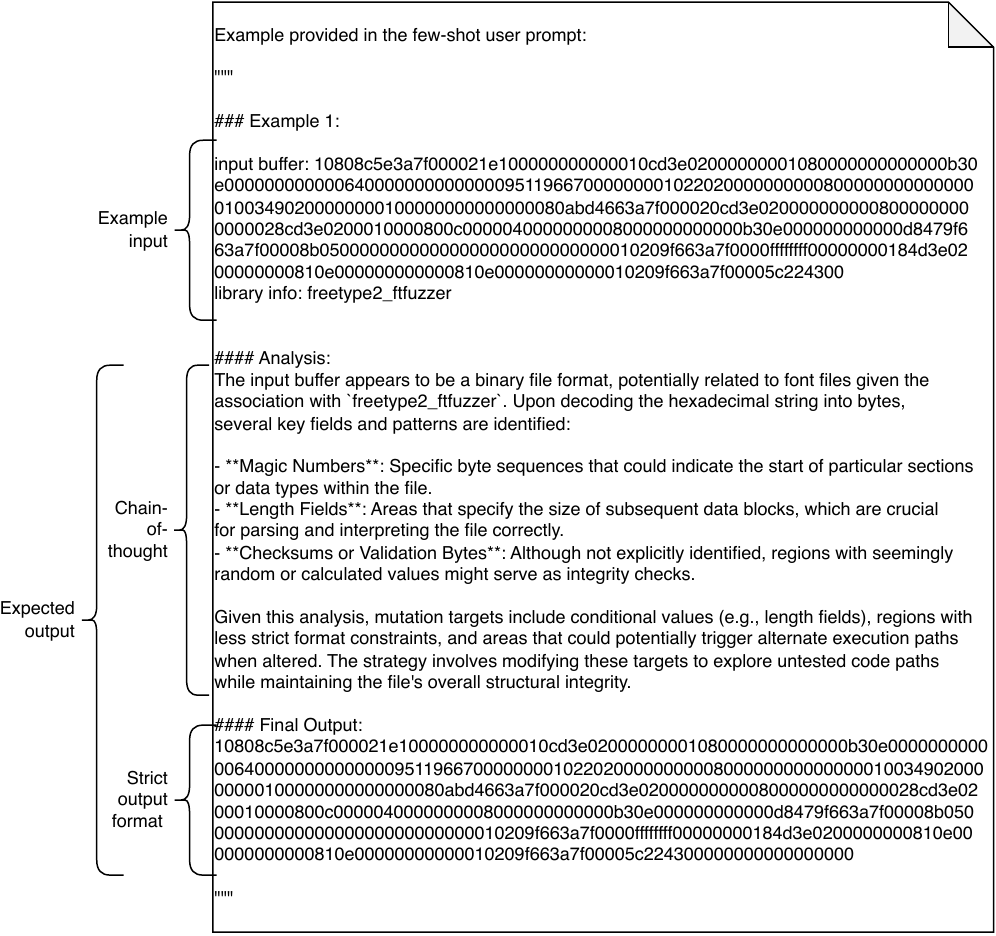}
        \caption{An in-context example provided in the user prompt.}
        \label{fig:ExamplePrompt}
    \end{figure}

User prompts supply dynamic, task-specific input data. However, due to the variability in LLM generated outputs \cite{guo2022survey}, strict formatting constraints alone may not ensure consistency. To improve stability, we apply few-shot prompting in the user prompt with embedded concrete examples that follow the required output format, and we also reinforce format reminders to minimize the chance of context drift. Figure \ref{fig:UserPrompt} shows a few-shot prompt example. Our research explores three prompt engineering strategies \cite{promptingguide}:
\squishlist
    \item \textbf{Zero-shot prompting} provides only task instructions without any examples or demonstrations. As shown in Figure \ref{fig:ZeroShotPrompt}, the “\textit{Example Output}” section does not contain actual mutated buffers. 
    \item \textbf{Few-shot prompting} requires explicit examples in the prompt. One-shot and three-shot prompting insert one and three successful mutation examples, respectively. These examples (Figure \ref{fig:ExamplePrompt}) are previous LLM outputs that met the formatting requirements.
    \item \textbf{Chain-of-thought prompting} \cite{cot} adds intermediate reasoning steps in prompts. Our few-shot prompting examples include an \emph{Analysis} section explaining the logic behind mutations—such as why certain characters are mutated in the input buffer, and how the final output is generated. To avoid context loss, the strict output format specification has to be restated after the examples.
\squishlistend
Effective prompt engineering often poses a challenge, as optimal prompts are rarely successful on the first attempt. Thus, the process typically requires iterative refinement. We adopt a continuous trial-and-error approach \cite{dang2022beyond}, which involves generating responses using the current prompt, evaluating whether the outputs meet structural and semantic requirements, making small targeted adjustments—such as emphasizing formatting constraints, altering example structures, or refining key terms—and repeating this process until the prompts consistently produce satisfactory results. 
This process produces the final versions of the zero-shot, one-shot and three-shot prompts, retained only after meeting strict output format requirements.

\section{Experimental Setup}
While fuzzers can theoretically run endlessly to discover bugs, in practical and industrial scenarios, execution time is limited. To reflect real-world constraints, we evaluated each configuration over two fixed time intervals: \emph{one hour} and \emph{four hours}. Each fuzzer and benchmark pair was tested using \emph{three independent trials} to reduce randomness and improve result reliability. All experiments were conducted on a high-performance server running \textit{Ubuntu 20.04.2 LTS}, equipped with an \textit{Intel Xeon Gold 5218 CPU (64 cores)}, \textit{754 GiB} of RAM, \textit{3 TB} of disk storage, and two \textit{NVIDIA Quadro RTX 6000 GPUs (24 GiB VRAM each)}. This section defines the performance metrics and describes the selected benchmarks, baseline, and LLMs. Extra experiment setup details and generated results are illustrated in the Section \ref{results}.

\subsection{Evaluation Metrics} \label{eval}
We use FuzzBench \cite{metzman2021fuzzbench} to collect and analyze coverage data across multiple targets, evaluating code coverage from four perspectives: (1) \emph{Function Coverage}, which measures the proportion of functions executed relative to the total in the executable; (2) \emph{Line Coverage}, assessing the percentage of executable lines exercised within functions; (3) \emph{Region Coverage}, evaluating distinct control-flow regions, which are sequences of instructions with a single entry and exit point; and (4) \emph{Branch Coverage}, also known as decision coverage, measuring executed branches in conditional structures—such as if, switch-case, loop, and try-catch statements—to ensure both true and false paths are explored.

We evaluate our LLM-guided mutation strategy primarily through code coverage, using \emph{Coverage Improvement Percentage (CIP)} to calculate LLM efficiency gains (\(\text{Coverage}_{\text{LLM}}\)) over a baseline fuzzer ( \(\text{Coverage}_{\text{Baseline}}\)), where positive values indicate the LLM outperforms the baseline.
The metric quantifies enhancements in fuzzing efficiency and implicitly reflects the LLM's mutation quality, as better mutations are likely to explore more code.
CIP is expressed as: 
\[
\text{CIP} (\%) = \text{Coverage}_{\text{LLM}}  (\%) -\text{Coverage}_{\text{Baseline}}  (\%)
\]

The \emph{Syntactic Correctness Rate (SCR)} evaluates the percentage of LLM outputs that follow the required format, reflecting how well prompts guide the model to produce usable results. $N_{correct}$ denotes the number of syntactically valid LLM-generated responses, and $N_{total}$ denotes the total number of LLM responses. SCR is expressed as:
\[
\text{SCR} = \frac{N_{\text{correct}}}{N_{\text{total}}} \times 100\%
\]
Despite strict prompt formatting and zero temperature for deterministic outputs, LLMs can still produce syntactic errors. These syntactic failures arise from two primary issues: \emph{hexadecimal conversion exceptions}—when a hex string fails to parse into a Python bytes object—and \emph{format mismatches}, where the required “Final Output” section is missing or contains invalid content. While we manually handle conversion exceptions, format mismatches remain the main cause of syntactic errors. Thus, SCR measures how often the LLM produces syntactically valid and properly formatted mutations. A higher SCR indicates more mutations with valid or usable formats, potentially enhancing fuzzing effectiveness.

LLM response stability can affect fuzzing effectiveness, as repeated outputs reduce mutation diversity and limit code coverage. Thus, we propose the Response Duplication Rate (RDR) as a metric for measuring LLM determinism.
Responses are duplicates if their “Final Output” matches a prior observed output within the same fuzzer-benchmark pair and prompt-shot setup. A high RDR signals low mutation diversity, suggesting that the LLM may be overly deterministic or insensitive to subtle prompt variations. Formally, RDR is the ratio of duplicated LLM responses ($N_\text{Duplicate}$) to the total number of LLM generations ($N_\text{Total}$):
\[
\text{RDR} = \frac{N_\text{Duplicate}}{N_\text{Total}} * 100\%
\]
In summary, we evaluates the proposed approach using \emph{four major metrics}: \emph{code coverage}, which measures the fuzzer's ability to explore program execution paths; \emph{CIP}, which qualifies performance gains relative to baseline fuzzers; \emph{SCR}, which assesses the reliability of LLM-generated mutations under structured prompt guidance; and \emph{RDR}, which evaluates how deterministic LLM-generated outputs are. 
\subsection{Selection of Benchmarks, Baselines, and LLMs}
Since our study focuses on exploring fuzzing in domains with high reliability demands and complex internal logic—namely, IoT firmware, mobile platforms, and autonomous driving systems—we selected 25 of the 26 benchmarks officially maintained in Fuzzbench \cite{metzman2021fuzzbench}. Their corresponding fuzz targets are derived from OSS-fuzz \cite{203944} and represent a diverse range of real-world open-source software programs \cite{metzman2021fuzzbench}. These benchmarks include structured input types such as file formats, network protocols, and cryptographic libraries, which closely match the input characteristics of our target domains and are compatible with our LLM mutation strategy. Within Fuzzbench, 20 benchmarks are \emph{general-purpose}, aiming to maximize code coverage across the target program, while 5 are \emph{specific benchmarks}, designed to reproduce known or suspected crashes.

Additionally, we select \emph{AFL} \cite{10.1145/3580596}, \emph{AFL++} \cite{10.5555/3488877.3488887}, and \emph{LibFuzzer} \cite{libfuzzerDocs} as baseline fuzzers due to their foundational roles and architectural relevance to our proposed system. These widely adopted, code coverage-guided grey-box fuzzers have influenced the development of modern fuzzing tools. Since our solution builds on AFL++, it is naturally selected as a baseline for direct comparison. AFL++, which extends AFL with hybrid mutation strategies and improved scheduling, involves AFL as a proper baseline to assess the evolution of fuzzing capabilities. LibFuzzer is also included as it is widely used in FuzzBench and serves as the foundation for several modern fuzzing tools (e.g. LibFuzzer-bin \cite{libfuzzbin}, LibAFL \cite{libafl}, and PromptFuzz \cite{10.1145/3658644.3670396}).
    
Effective LLM-guided fuzzing requires models with four core capabilities: (1) strong \emph{code generation ability} to produce structured, syntactically valid mutations; (2) solid \emph{language understanding} to recognize input formats and maintain structural integrity; (3) reliable \emph{instruction following} to strictly fulfill prompt requirements; and (4) effective \emph{reasoning} to explain mutation logic and decision-making. Based on these criteria, we select Llama3.3 (70B), Deepseek-r1-Distill-Llama-70B, QwQ-32B, and Gemma3-27B as baseline models. These open-source models represent recent advances in code-centric LLMs and excel in structured text generation, instruction compliance, and semantic reasoning. They are also compatible with our system's resource constraints.

\section{Results and Discussions} \label{results}
To assess the potential of state-of-the-art reasoning-capable LLMs in enhancing fuzzing performance, we developed a dedicated infrastructure that integrates traditional fuzzing engines with LLM-driven mutation components. This infrastructure serves as the foundation for all experimental evaluations addressing our research questions. To answer \textbf{R2}, we conducted a focused set of experiments evaluating the impact of prompt engineering—specifically 0-shot, 1-shot, and 3-shot configurations—on the quality of mutations generated by Llama3.3 \cite{2024arXiv240721783G}. To investigate whether other LLMs outperform Llama3.3, addressing \textbf{R3} and \textbf{R4}, we further experimented and compared code coverage achieved by applying various state-of-the-art reasoning models under different prompt shots. The selected LLMs are widely recognized for their strong performance in the AI community and represent current advances in reasoning capabilities. 

We also evaluate results across four metrics—code coverage, CIP, SCR, and RDR—introduced in Section \ref{eval}, from three primary perspectives: code coverage, mutation quality, and fuzzing efficiency. Code coverage data were extracted from FuzzBench-generated reports, including function, line, branch, and region coverage. Additionally, approximately \emph{96GiB} of logs were generated by the LLM component. Through data cleaning and log parsing, we constructed \emph{log analysis tables} containing over \textit{$640,000$} entries.

\subsection{Baseline Comparison Experiment}
To establish a baseline for comparison, each of the three mutation-based fuzzers—\emph{AFL++} \cite{10.5555/3488877.3488887}, \emph{AFL} \cite{10.1145/3580596}, and \emph{LibFuzzer} \cite{libfuzzerDocs}—was evaluated across \emph{three independent trials} using the FuzzBench framework. Each fuzzer was tested against a common set of \emph{25 selected benchmarks} per trial.
All fuzzers were run with their \emph{default configurations} defined in FuzzBench, which apply basic mutation strategies such as \emph{Havoc} and \emph{MOpt} \cite{mopt}. 

\begin{table}[]
\centering
\caption{The number of benchmarks that has the maximum code coverage results for each baseline fuzzer. \label{long}}
\label{table:base-count}
\begin{tabular}{lllll}
\hline \hline
\multicolumn{5}{c}{\textbf{ Frequencies of Maximum Code Coverage Occurred for Baseline}}                                                         \\ \hline \hline
\multicolumn{1}{l|}{\textbf{RunTime}}   & \multicolumn{1}{l|}{\textbf{Coverage-type}} & \textbf{AFL++} & \textbf{AFL} & \textbf{LibFuzzer} \\ \hline
\multicolumn{1}{l|}{\multirow{4}{*}{1}} & \multicolumn{1}{l|}{Branch Coverage}        & \textbf{18}             & 12           & 10                 \\
\multicolumn{1}{l|}{}                   & \multicolumn{1}{l|}{Function Coverage}      & \textbf{17}   & 7            & 5                  \\
\multicolumn{1}{l|}{}                   & \multicolumn{1}{l|}{Line Coverage}          & \textbf{16}             & 8            & 3                  \\
\multicolumn{1}{l|}{}                   & \multicolumn{1}{l|}{Region Coverage}        & \textbf{18}             & 9            & 4                  \\ \hline
\multicolumn{1}{l|}{\multirow{4}{*}{4}} & \multicolumn{1}{l|}{Branch Coverage}        & \textbf{17}             & 10           & 14                 \\
\multicolumn{1}{l|}{}                   & \multicolumn{1}{l|}{Function Coverage}      & \textbf{18}             & 4            & 7                  \\
\multicolumn{1}{l|}{}                   & \multicolumn{1}{l|}{Line Coverage}          & \textbf{13}             & 5            & 9                  \\
\multicolumn{1}{l|}{}                   & \multicolumn{1}{l|}{Region Coverage}        & \textbf{18}             & 5            & 6                  \\ \hline \hline
\end{tabular}
\end{table}

Based on one-hour and four-hour testing results, \emph{Table \ref{table:base-count}} summarizes the frequency of highest code coverage achieved by each baseline fuzzer. \emph{AFL++ consistently outperformed} other fuzzers across all coverage types and time intervals. For instance, it achieved the highest branch coverage on \textit{18 benchmarks} in the one-hour runs, which is over \emph{$69\%$} of all baseline fuzzer-benchmark pairs. In the four-hour runs, AFL++ outperformed both AFL and LibFuzzer on \textit{17 benchmarks} (\emph{$65\%$}). Given consistently superior performance of AFL++ across both time intervals, \emph{AFL++ is used as the primary baseline} for evaluating the effectiveness of LLM-guided fuzzers in subsequent experiments.

\subsection{Llama3.3 Prompt Engineering Evaluation Experiment} \label{prompt-shot-exp}
This experiment investigates how prompt engineering— specifically the number of prompt shots—impacts the effectiveness of LLM-guided fuzzing, thereby addressing \textbf{R2}. 
Since prompt design is inherently iterative and costly to refine within the full FuzzBench pipeline, we initially optimized prompt templates by using \emph{Llama3.3} model in a standalone \emph{Python} environment. 
Finalized prompts for 0-shot, 1-shot, and 3-shot settings are presented in \emph{Section \ref{prompt-eng}: Prompt Refinement}.
After finalizing prompts, we set the LLM's temperature to $0.0$ for deterministic and reproducible outputs, and loaded the full-scale \emph{Llama3.3-70B model} (approximately \emph{43 GB of VRAM}) into the Fuzzbench pipeline with these templates and settings. For each prompt shot, the system was initialized from scratch, and logs for all trials were preserved for reproducibility, traceability and analysis.

\emph{Table \ref{table:ps-branch}} highlights trends in branch coverage percentages for 0-shot, 1-shot, and 3-shot prompts from both one-hour and four-hour fuzzing tests using Llama3.3. 
The effect of increasing prompt shots on coverage is observed to be \emph{highly benchmark-dependent}: \emph{libpcap\_fuzz\_both} and \emph{vorbis\_decode\_fuzzer} benchmarks show improved branch coverage with increased shots, while \emph{harfbuzz\_hb-shape-fuzzer} and \emph{sqlite3\_ossfuzz} 
show little or no gain, and \emph{libpng\_libpng\_read\_fuzzer} and \emph{openh264\_decoder\_fuzzer} even exhibit reduced coverage in four-hour runs.
Moreover, analysis of LLM response logs revealed that Ollama response timeouts occurred in approximately \emph{35\%} of cases. Although these timeouts may reduce usable mutations, response quality is separately assessed using SCR and RDR metrics. The following sections plot the relationship between LLM response quality and code coverage, offering insight into the instability caused by varying prompt shots. 

\begin{table*}[]
\centering
\caption{Branch coverage for multiple prompt shots in Llama3.3.\label{long}}
\label{table:ps-branch}
\begin{tabular}{lllllllll}
\hline \hline
\multicolumn{9}{c}{\textbf{Branch Coverage: Llama3.3 Prompt Engineering Experiment}}              \\ \hline \hline
\multicolumn{1}{l|}{\multirow{2}{*}{\textbf{Benchmarks}}}                      & \multicolumn{4}{l|}{\textbf{One-Hour Runtime}}                                                                                                                                                                                                                                           & \multicolumn{4}{l}{\textbf{Four-Hour Runtime}}                                                                                                                                                                                                                      \\ \cline{2-9}
\multicolumn{1}{l|}{} 
                                                           & \textbf{\begin{tabular}[c]{@{}l@{}}0-shot \\ (\%)\end{tabular}} & \textbf{\begin{tabular}[c]{@{}l@{}}1-shot \\ (\%)\end{tabular}} & \textbf{\begin{tabular}[c]{@{}l@{}}3-shot \\ (\%)\end{tabular}} & \multicolumn{1}{l|}{\textbf{\begin{tabular}[c]{@{}l@{}}AFL++\\ (\%)\end{tabular}}} & \textbf{\begin{tabular}[c]{@{}l@{}}0-shot \\ (\%)\end{tabular}} & \textbf{\begin{tabular}[c]{@{}l@{}}1-shot \\ (\%)\end{tabular}} & \textbf{\begin{tabular}[c]{@{}l@{}}3-shot \\ (\%)\end{tabular}} & \textbf{\begin{tabular}[c]{@{}l@{}}AFL++\\ (\%)\end{tabular}} \\ \hline
\multicolumn{1}{l|}{freetype2\_ftfuzzer}                   & \textbf{37.66}                                                  & 37.27                                                           & 37.29                                                           & \multicolumn{1}{l|}{36.94}                                                         & 39.47                                                           & 40.01                                                           & \textbf{40.26}                                                  & 38.72                                                         \\
\multicolumn{1}{l|}{libxml2\_xml}                          & 13.59                                                           & \textbf{22.17}                                                  & 13.59                                                           & \multicolumn{1}{l|}{23.88}                                                         & 13.42                                                           & \textbf{13.59}                                                  & \textbf{13.59}                                                  & 24.77                                                         \\
\multicolumn{1}{l|}{libpng\_libpng\_read\_fuzzer}          & 32.72                                                           & \textbf{32.95}                                                  & \textbf{32.95}                                                  & \multicolumn{1}{l|}{33.17}                                                         & \textbf{32.90}                                                  & 32.77                                                           & 32.80                                                           & 33.57                                                         \\
\multicolumn{1}{l|}{bloaty\_fuzz\_target}                  & 5.73                                                            & 5.73                                                            & \textbf{6.74}                                                   & \multicolumn{1}{l|}{7.20}                                                          & \textbf{6.22}                                                   & 6.20                                                            & 5.73                                                            & 7.69                                                          \\
\multicolumn{1}{l|}{curl\_curl\_fuzzer\_http}              & 12.72                                                           & \textbf{12.78}                                                  & 12.65                                                           & \multicolumn{1}{l|}{13.07}                                                         & \textbf{12.69}                                                  & 12.66                                                           & 12.67                                                           & 13.35                                                         \\
\multicolumn{1}{l|}{harfbuzz\_hb-shape-fuzzer}             & \textbf{44.21}                                                  & \textbf{44.21}                                                  & \textbf{44.21}                                                  & \multicolumn{1}{l|}{61.17}                                                         & \textbf{44.21}                                                  & \textbf{44.21}                                                  & \textbf{44.21}                                                  & 63.44                                                         \\
\multicolumn{1}{l|}{jsoncpp\_jsoncpp\_fuzzer}              & \textbf{24.41}                                                  & \textbf{24.41}                                                  & 0.42                                                            & \multicolumn{1}{l|}{24.46}                                                         & \textbf{0.42}                                                   & \textbf{0.42}                                                   & \textbf{0.42}                                                   & 24.46                                                         \\
\multicolumn{1}{l|}{lcms\_cms\_transform\_fuzzer}          & 0.50                                                            & 0.50                                                            & \textbf{20.26}                                                  & \multicolumn{1}{l|}{23.56}                                                         & 23.07                                                           & \textbf{25.34}                                                  & 22.83                                                           & 19.39                                                         \\
\multicolumn{1}{l|}{libjpeg-turbo\_libjpeg\_turbo\_fuzzer} & \textbf{26.39}                                                  & \textbf{26.39}                                                  & \textbf{26.39}                                                  & \multicolumn{1}{l|}{26.69}                                                         & \textbf{26.39}                                                  & \textbf{26.39}                                                  & \textbf{26.39}                                                  & 26.70                                                         \\
\multicolumn{1}{l|}{libpcap\_fuzz\_both}                   & 36.48                                                           & 36.51                                                           & \textbf{37.91}                                                  & \multicolumn{1}{l|}{37.83}                                                         & \textbf{35.66}                                                  & 34.16                                                           & 34.67                                                           & 40.24                                                         \\
\multicolumn{1}{l|}{mbedtls\_fuzz\_dtlsclient}             & \textbf{9.84}                                                   & \textbf{9.84}                                                   & \textbf{9.84}                                                   & \multicolumn{1}{l|}{14.16}                                                         & 13.03                                                           & \textbf{13.44}                                                  & 9.84                                                            & 14.42                                                         \\
\multicolumn{1}{l|}{openh264\_decoder\_fuzzer}             & 74.53                                                           & 74.34                                                           & \textbf{74.55}                                                  & \multicolumn{1}{l|}{74.52}                                                         & \textbf{76.40}                                                  & 76.00                                                           & 75.42                                                           & 76.10                                                         \\
\multicolumn{1}{l|}{openssl\_x509}                         & \textbf{10.72}                                                  & \textbf{10.72}                                                  & \textbf{10.72}                                                  & \multicolumn{1}{l|}{10.80}                                                         & \textbf{10.72}                                                  & \textbf{10.72}                                                  & \textbf{10.72}                                                  & 10.83                                                         \\
\multicolumn{1}{l|}{openthread\_ot-ip6-send-fuzzer}        & 11.20                                                           & 10.29                                                           & \textbf{11.43}                                                  & \multicolumn{1}{l|}{11.41}                                                         & \textbf{11.23}                                                  & 10.94                                                           & 11.20                                                           & 11.50                                                         \\
\multicolumn{1}{l|}{proj4\_proj\_crs\_to\_crs\_fuzzer}     & 8.81                                                            & 8.41                                                            & \textbf{9.05}                                                   & \multicolumn{1}{l|}{9.11}                                                          & \textbf{10.60}                                                  & 10.35                                                           & 10.50                                                           & 10.73                                                         \\
\multicolumn{1}{l|}{re2\_fuzzer}                           & 62.89                                                           & 62.55                                                           & \textbf{63.00}                                                  & \multicolumn{1}{l|}{64.27}                                                         & 64.02                                                           & 63.85                                                           & \textbf{64.07}                                                  & 64.34                                                         \\
\multicolumn{1}{l|}{sqlite3\_ossfuzz}                      & \textbf{26.91}                                                  & \textbf{26.91}                                                  & \textbf{26.91}                                                  & \multicolumn{1}{l|}{45.58}                                                         & \textbf{26.91}                                                  & \textbf{26.91}                                                  & \textbf{26.91}                                                  & 63.30                                                         \\
\multicolumn{1}{l|}{stb\_stbi\_read\_fuzzer}               & \textbf{55.08}                                                  & \textbf{55.08}                                                  & \textbf{55.08}                                                  & \multicolumn{1}{l|}{68.51}                                                         & \textbf{55.08}                                                  & \textbf{55.08}                                                  & \textbf{55.08}                                                  & 69.40                                                         \\
\multicolumn{1}{l|}{systemd\_fuzz-link-parser}             & \textbf{22.90}                                                  & 21.90                                                           & 22.10                                                           & \multicolumn{1}{l|}{22.30}                                                         & 21.90                                                           & \textbf{22.30}                                                  & 22.20                                                           & 22.30                                                         \\
\multicolumn{1}{l|}{vorbis\_decode\_fuzzer}                & 30.47                                                           & \textbf{30.49}                                                  & 30.47                                                           & \multicolumn{1}{l|}{30.86}                                                         & 30.54                                                           & 30.54                                                           & \textbf{30.59}                                                  & 30.83                                                         \\
\multicolumn{1}{l|}{woff2\_convert\_woff2ttf\_fuzzer}      & \textbf{17.02}                                                  & \textbf{17.02}                                                  & \textbf{17.02}                                                  & \multicolumn{1}{l|}{26.05}                                                         & \textbf{17.02}                                                  & \textbf{17.02}                                                  & \textbf{17.02}                                                  & 26.74                                                         \\ \hline
\multicolumn{1}{l|}{php\_php-fuzz-parser\_0dbedb}          & \textbf{9.67}                                                   & 9.62                                                            & 9.62                                                            & \multicolumn{1}{l|}{9.74}                                                          & \textbf{9.93}                                                   & 9.86                                                            & 9.87                                                            & 9.90                                                          \\
\multicolumn{1}{l|}{mbedtls\_fuzz\_dtlsclient\_7c6b0e}     & 13.69                                                           & \textbf{13.73}                                                  & 13.58                                                           & \multicolumn{1}{l|}{13.48}                                                         & 13.69                                                           & 13.74                                                           & \textbf{13.86}                                                  & 13.67                                                         \\
\multicolumn{1}{l|}{libxml2\_xml\_e85b9b}                  & \textbf{23.06}                                                  & 20.97                                                           & 20.99                                                           & \multicolumn{1}{l|}{23.49}                                                         & 20.18                                                           & 13.42                                                           & \textbf{23.14}                                                  & 28.53                                                         \\
\multicolumn{1}{l|}{harfbuzz\_hb-shape-fuzzer\_17863b}     & \textbf{45.02}                                                  & \textbf{45.02}                                                  & \textbf{45.02}                                                  & \multicolumn{1}{l|}{55.09}                                                         & \textbf{45.02}                                                  & \textbf{45.02}                                                  & \textbf{45.02}                                                  & 59.77                                                         \\
\multicolumn{1}{l|}{bloaty\_fuzz\_target\_52948c}          & 5.73                                                            & \textbf{6.11}                                                   & 6.06                                                            & \multicolumn{1}{l|}{6.55}                                                          & 6.33                                                            & 5.84                                                            & \textbf{6.44}                                                   & 7.38                                                          \\
\hline \hline
\end{tabular}
\end{table*}

\subsubsection{Plot Analysis: SCR and RDR Across Benchmarks}
\begin{figure}[t]
    \centering
    \includegraphics[width=0.45\textwidth]{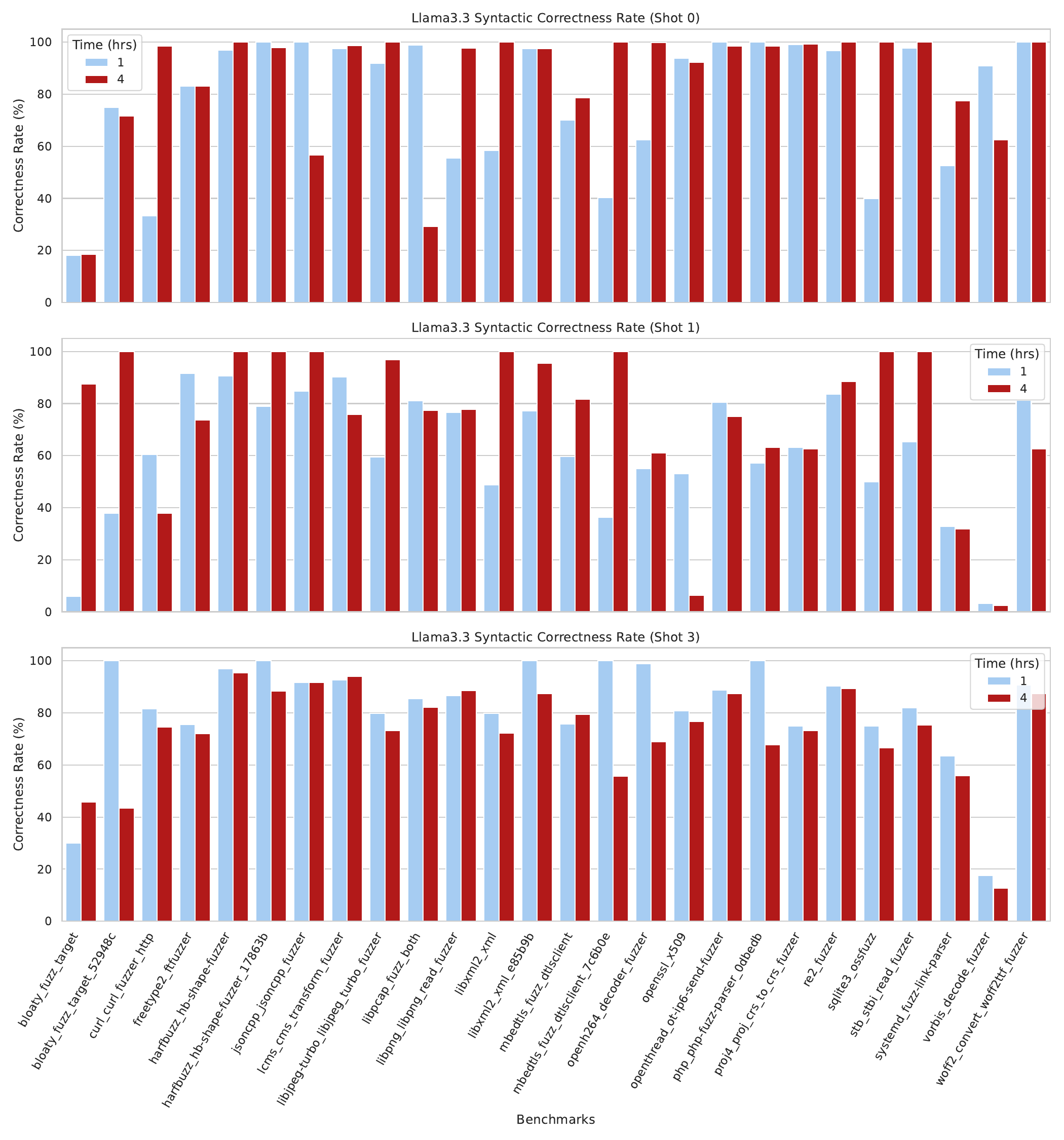}
    \caption{Llama3.3 SCR for prompt shots 0, 1, and 3 against selected benchmarks}
    \label{Fig:llama-scr}
\end{figure}

\begin{figure}[t]
    \centering
    \includegraphics[width=0.45\textwidth]{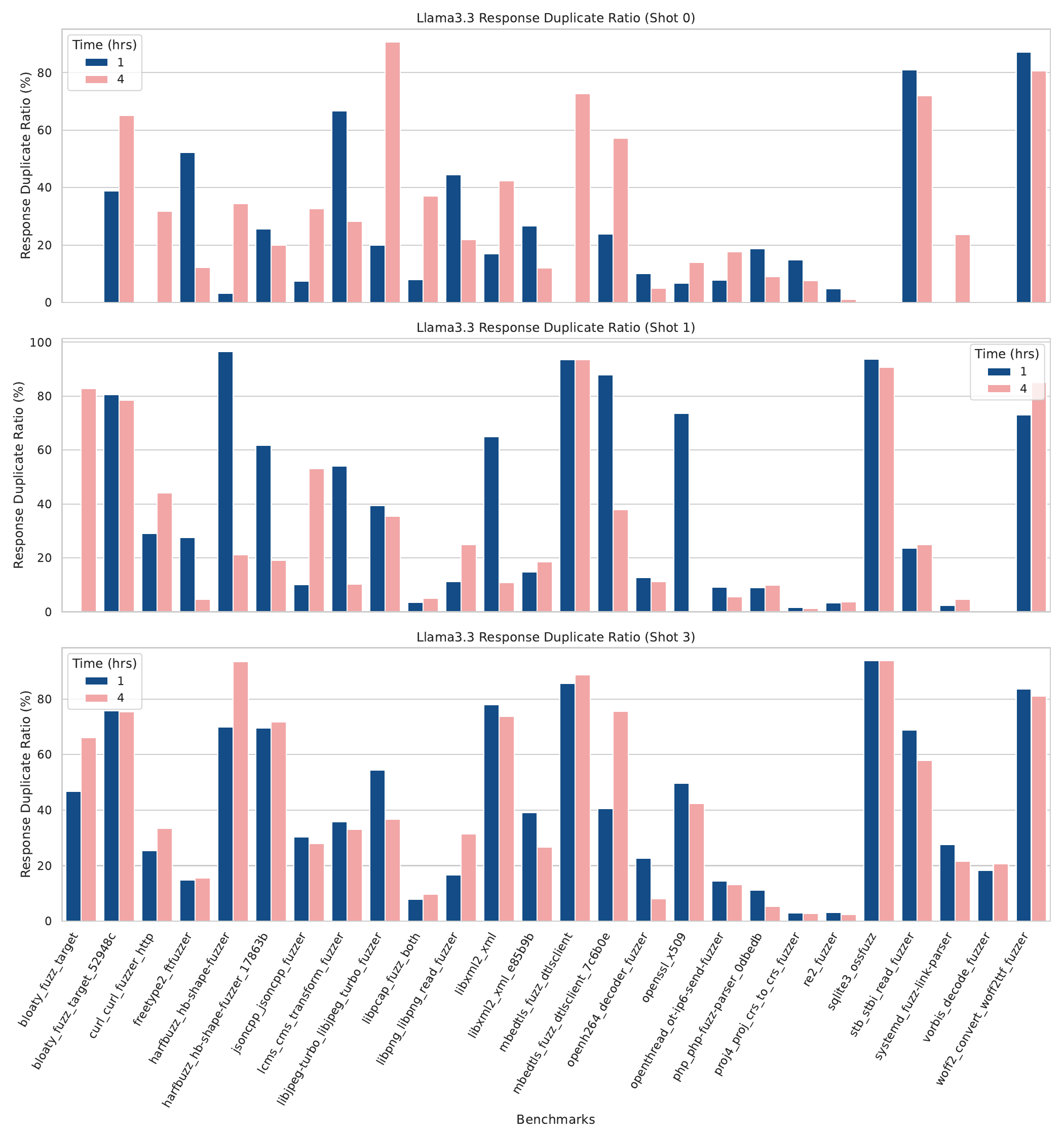}
    \caption{Llama3.3 Response Duplicate Ratio (RDR) for prompt shots 0, 1, and 3 against selected benchmarks}
    \label{Fig:llama-rdr}
\end{figure}

We compute both metrics based on logs of each fuzzer–benchmark pair execution, using structured \emph{log analysis tables}. Syntactic-valid responses are counted from recorded LLM-generated outputs in the tables, while duplicates are detected by comparing the “Final Output” sections within the same prompt shot and fuzzer-benchmark pair.
\emph{Figures \ref{Fig:llama-scr}} and \emph{\ref{Fig:llama-rdr}} present SCR and RDR for each benchmark across different prompt shots and time intervals, illustrating response validity and duplication tendencies. Each benchmark in the plots relates to two fuzzing durations—one-hour and four-hour runs—with colors distinguishing these intervals for ease of comparison.

Examining the SCR plots, we observe that SCR rates slightly drop for most benchmarks as prompt shots increase, hinting that lengthy or complex prompts may raise the risk of formatting errors. Benchmarks like openthread\_ot-ip6-send-fuzzer and woff2\_convert\_woff2ttf\_fuzzer
maintain consistently high SCR values; while vorbis\_decode\_fuzzer shows \emph{notable decreases}.
In the RDR plots, benchmarks such as re2\_fuzzer and openthread\_ot-ip6-send-fuzzer remain at low values across all prompt shots, suggesting effective response diversification; whereas mbedtls\_fuzz\_dtlsclient and woff2\_convert\_woff2ttf\_fuzzer show high RDR values, indicating frequent repetition in LLM outputs that may limit the LLM's ability to explore new execution paths through diverse mutations. 
These observations lead to the question of whether formatting issues or response duplication hinder the mutation process and, in turn, the code coverage achieved by the fuzzer. To explore the relationship between code coverage and the syntactic correctness and diversity of LLM-generated responses, we generated additional visualizations correlating various code coverage types (function, line, branch, and region) with SCR and RDR across different fuzzing time and prompt shots. 
These results are presented and discussed in the following sections.

\subsubsection{Plot Analysis: Coverage vs. SCR} 
\begin{figure}[t]
    \centering
    \includegraphics[width=0.45\textwidth]{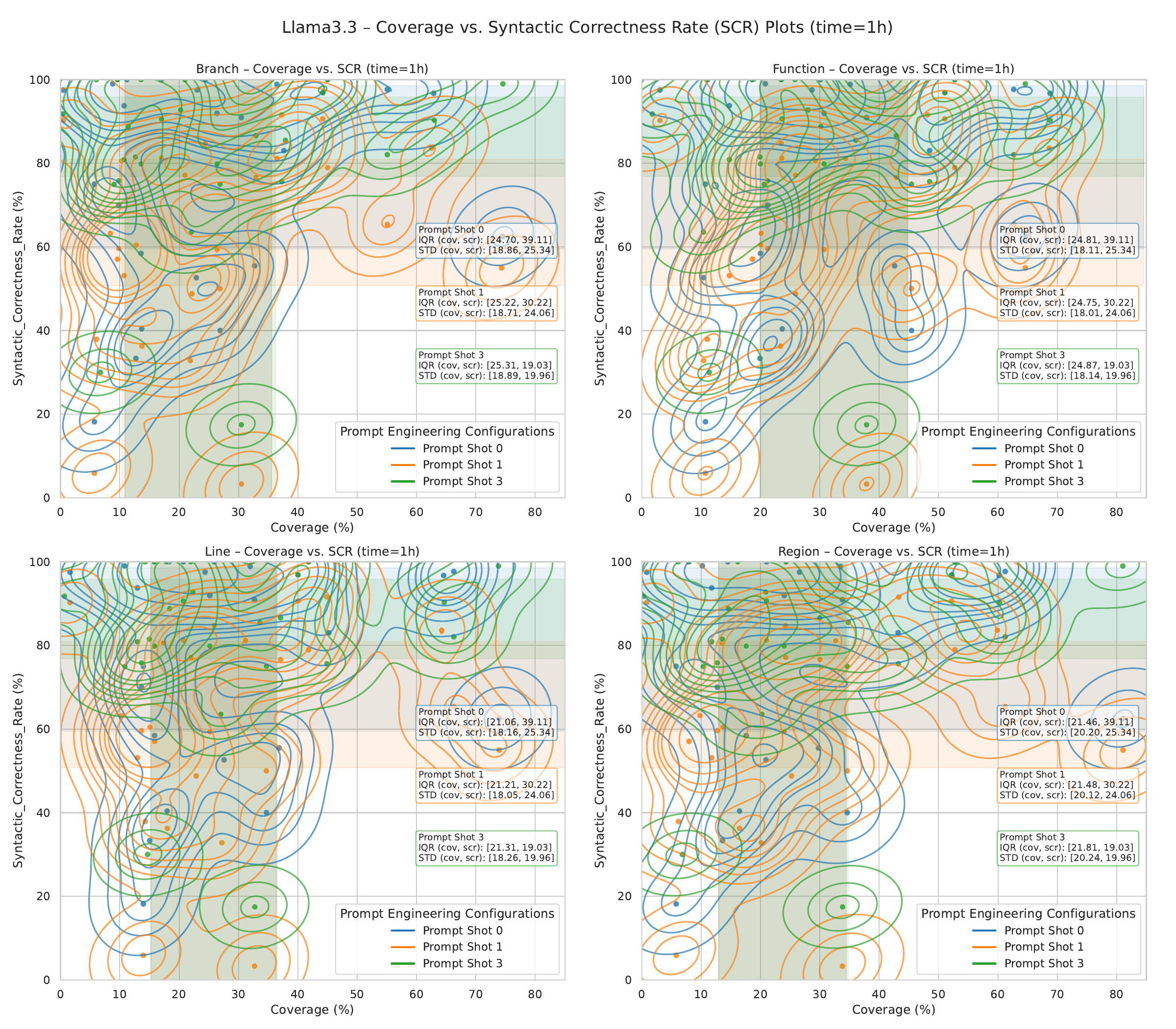}
    \caption{Llama3.3 Coverage vs SCR plots for prompt shots 0, 1, and 3 at one-hour time interval}
    \label{Fig:llama-1-cs-kde}
\end{figure}

\begin{figure}[t]
    \centering
    \includegraphics[width=0.45\textwidth]{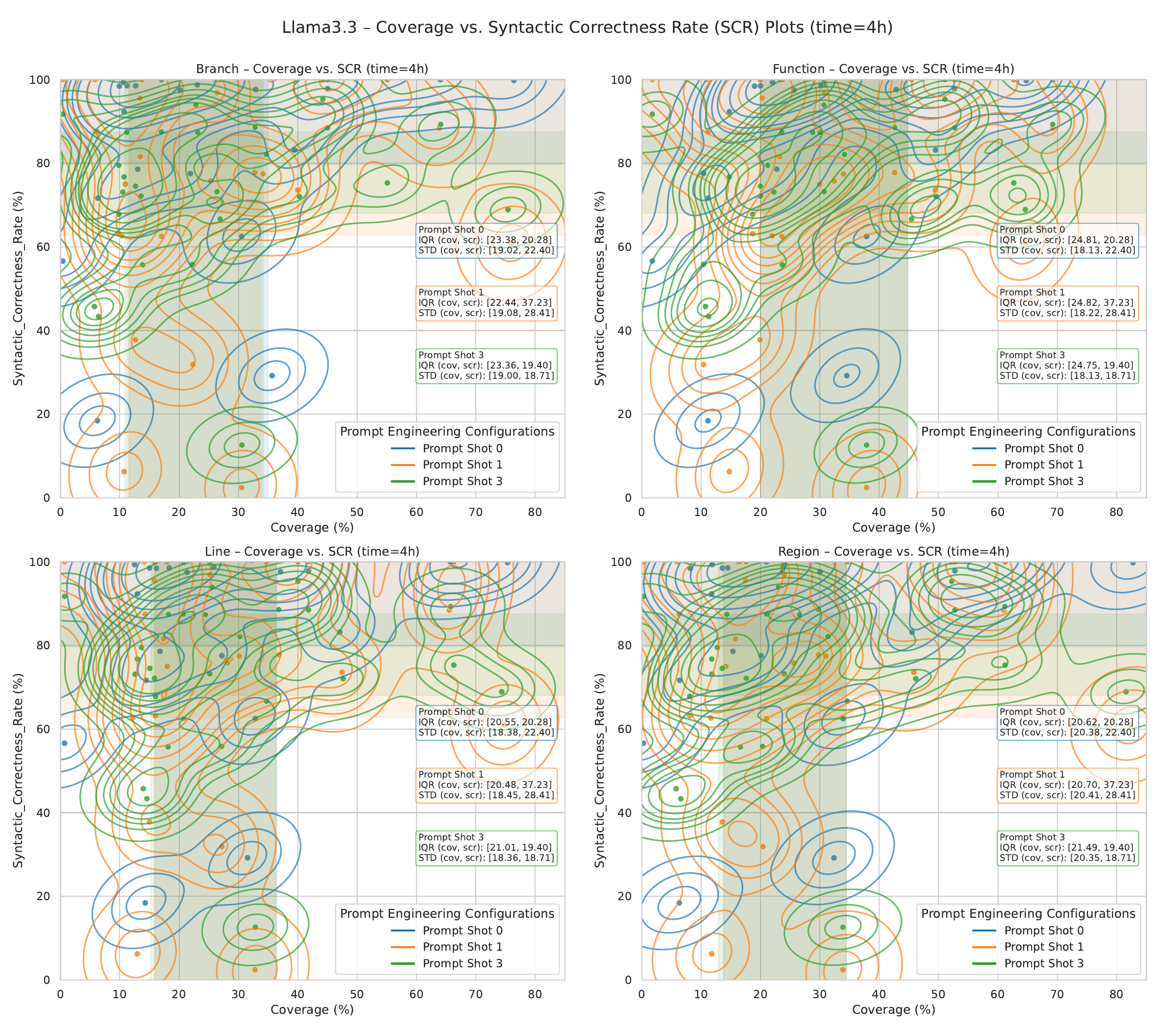}
    \caption{Llama3.3 Coverage vs SCR plots for prompt shots 0, 1, and 3 at four-hour time interval}
    \label{Fig:llama-4-cs-kde}
\end{figure}

\emph{Figures \ref{Fig:llama-1-cs-kde} and \ref{Fig:llama-4-cs-kde}} present the relationship between SCR and various code coverage metrics as prompt shots change over time. Each figure combines scatter and \emph{Kernel Density Estimate (KDE)} plots to highlight trends and correlations. Each data point represents a benchmark–fuzzer pair, with different prompt shots indicated by distinct colors—blue, orange, and green. The colored bands show the interquartile range (IQR) between the 0.25 and 0.75 quantiles, highlighting where most data points lie for each prompt shot.
The plots reveal that SCR values generally \emph{decrease with increased prompt shots}. Although some outliers exist, 3-shot prompts exhibit a more concentrated and predictable SCR distribution, with slightly narrower spreads and more centralized coverage than 0- and 1-shot prompts. This suggests that additional examples help stabilize LLM formatting. 
Across all prompt shots, SCR values cluster around \emph{70–80\%}, reflecting an "L-shaped" pattern in the plots: code coverage rises with SCR initially but plateaus beyond this range. 

\subsubsection{Plot Analysis: Coverage vs. RDR}
\begin{figure}[t]
    \centering
    \includegraphics[width=0.45\textwidth]{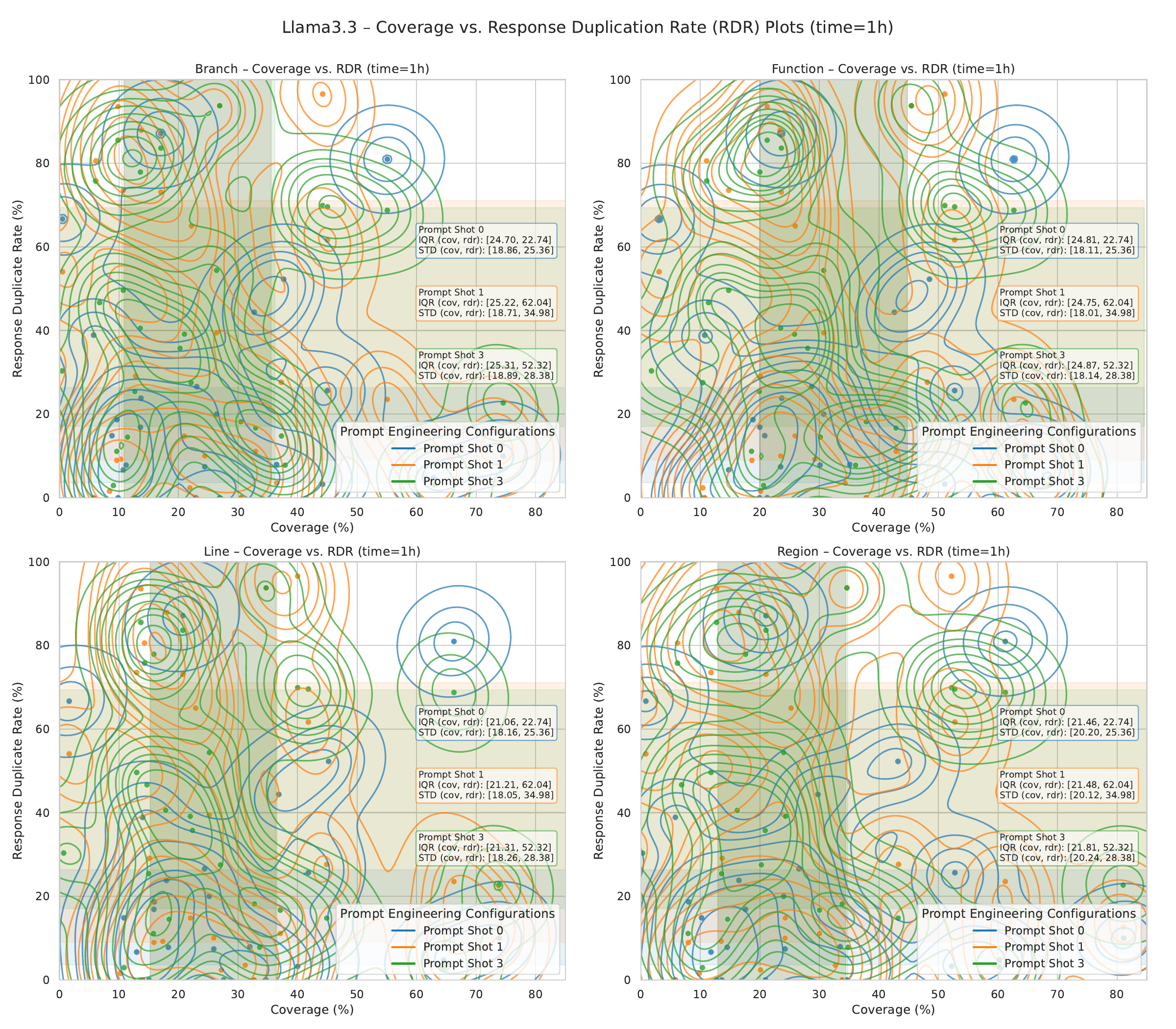}
    \caption{Llama3.3 Coverage vs RDR plots for prompt shots 0, 1, and 3 at one-hour time interval}
    \label{Fig:llama-1-cr-kde}
\end{figure}

\begin{figure}[t]
    \centering
    \includegraphics[width=0.45\textwidth]{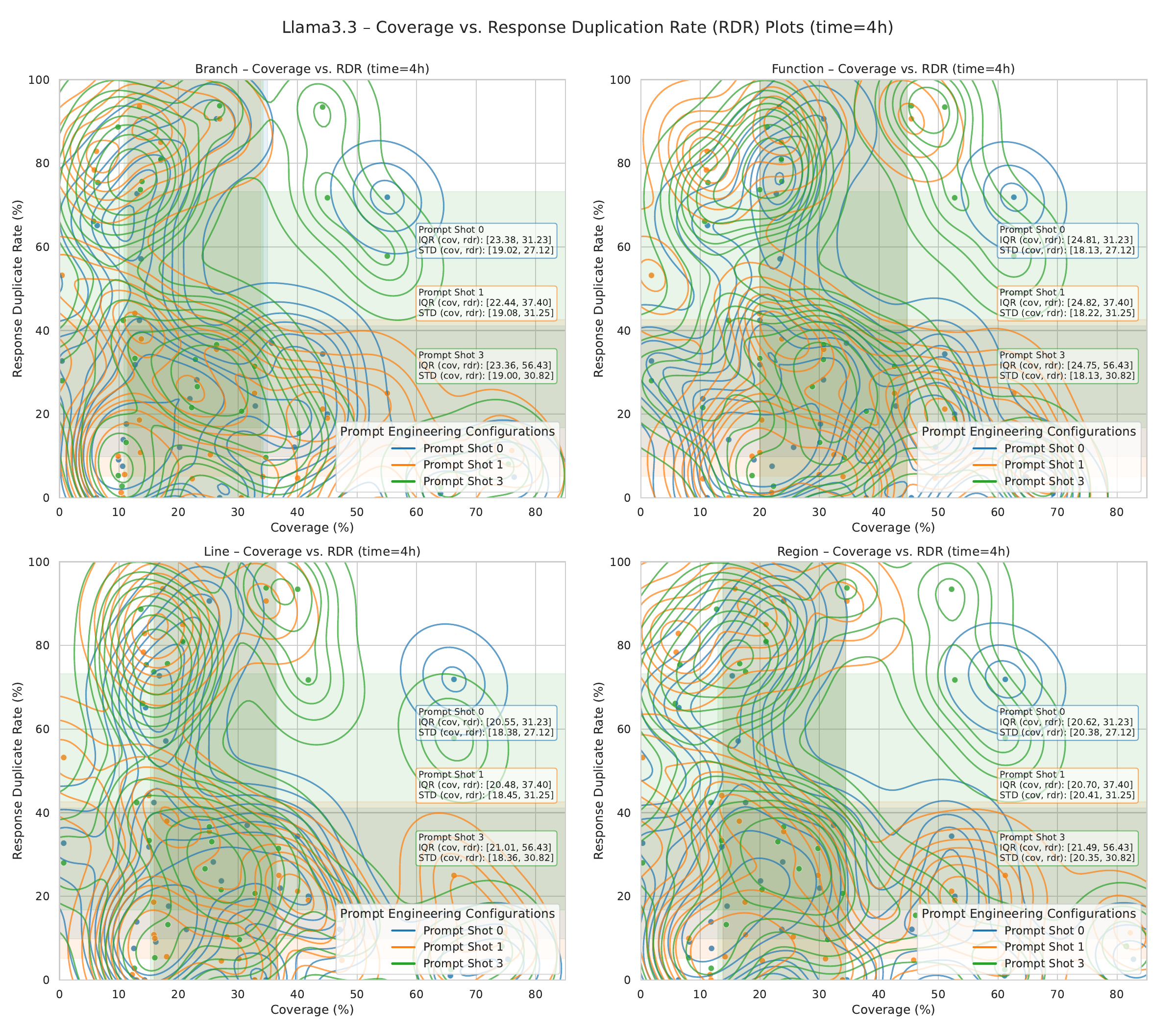}
    \caption{Llama3.3 Coverage vs RDR plots for prompt shots 0, 1, and 3 at four-hour time interval}
    \label{Fig:llama-4-cr-kde}
\end{figure}

Beyond syntactic correctness and conversion stability, the diversity of LLM-generated mutations plays a vital role in effective fuzzing. Repetitive or identical LLM outputs may reduce mutation variety and can limit code path exploration, even if syntactically valid.
To assess this factor, \emph{Figures \ref{Fig:llama-1-cr-kde} and \ref{Fig:llama-4-cr-kde}} present scatter and KDE plots relating RDR to final code coverage across various prompt shots and time intervals, where plot colors correspond to the prompt shots.

The plots show that lower RDRs generally correspond to higher code coverage, suggesting that \emph{greater response diversity promotes more effective fuzzing}; whereas higher RDRs cluster at lower coverage, indicating that repeated responses \emph{limit} novel input generation. This trend is more evident with the four-hour fuzzing interval. KDE plots reinforce this interpretation: \emph{3-shot prompts} generate more duplicate responses overall, retaining broader coverage and higher densities in lower duplication regions early on, but shifting toward narrower coverage and higher RDRs in longer runs. These observations imply that increasing prompt shots can generate more effective mutated inputs only when LLM response duplication remains low.

\subsection{Multiple LLMs Comparison Experiment} \label{multi-llm-exp}
This experiment evaluates the relative efficiency of multiple state-of-the-art LLMs under our mutation strategies, thereby addressing \textbf{R3} and \textbf{R4}. Based on the results of the \emph{Section \ref{prompt-shot-exp}}, we selected \emph{six benchmarks} with high code coverage variance, along with LLMs— \emph{Llama3.3} \cite{2024arXiv240721783G}, \emph{Deepseek-r1-Distill-Llama-70B} \cite{bau2024gradient}, \emph{QwQ-32B} \cite{qwen32b2024}, and \emph{Gemma3-27B} \cite{gemma3}—for this test. Since Llama3.3 had already been evaluated on these benchmarks in earlier experiments, it was excluded from execution but retained as a reference. The experimental setup mirrors that of the \emph{Section \ref{prompt-shot-exp}} experiment. Each LLM was evaluated in a fresh fuzzing session, loaded alone in Ollama, and tested across all six benchmarks, three prompt shots, and two time intervals. After each run, final code coverage was recorded. 
Based on these records, Table \ref{table:llm-count} summarizes how often each LLM achieved the highest coverage at different runtimes. The results indicate that \emph{Llama3.3} consistently outperforms other models during short (i.e., one hour) runs, whereas \emph{Deepseek-r1-Distill-Llama-70B} performs best mostly in four-hour runs.
\begin{table}[]
\centering
\caption{The number of benchmarks that has the maximum code coverage results for each LLM fuzzer. \label{long}}
\label{table:llm-count}
\begin{tabular}{lllll}
\hline \hline
\multicolumn{5}{l}{\textbf{Counts of Maximum Code Coverage for LLM Fuzzers}}                                                                                                                                                          \\ \hline \hline
\multicolumn{5}{c}{\textbf{RunTime = 1 Hour}}                                                                                                                                                                                                                         \\ \hline
\multicolumn{1}{l|}{\textbf{Coverage-type}} & \textbf{\begin{tabular}[c]{@{}l@{}}Llama\\ v3.3\end{tabular}} & \textbf{\begin{tabular}[c]{@{}l@{}}Deepseek\\ -r1\end{tabular}} & \textbf{\begin{tabular}[c]{@{}l@{}}QwQ\\ -32B\end{tabular}} & \textbf{Gemma3} \\ \hline
\multicolumn{1}{l|}{Branch Coverage}        & \textbf{7}                                                    & 3                                                               & 4                                                           & 4               \\
\multicolumn{1}{l|}{Function Coverage}      & \textbf{9}                                                    & 6                                                               & 6                                                           & 4               \\
\multicolumn{1}{l|}{Line Coverage}          & \textbf{8}                                                    & 2                                                               & 5                                                           & 3               \\
\multicolumn{1}{l|}{Region Coverage}        & \textbf{10}                                                   & 2                                                               & 4                                                           & 3               \\ \hline
\multicolumn{5}{c}{\textbf{RunTime = 4 Hour}}                                                                                                                                                                                                                 \\ \hline
\multicolumn{1}{l|}{\textbf{Coverage-type}} & \textbf{\begin{tabular}[c]{@{}l@{}}Llama\\ v3.3\end{tabular}} & \textbf{\begin{tabular}[c]{@{}l@{}}Deepseek\\ -r1\end{tabular}} & \textbf{\begin{tabular}[c]{@{}l@{}}QwQ\\ -32B\end{tabular}} & \textbf{Gemma3} \\ \hline
\multicolumn{1}{l|}{Branch Coverage}        & 6                                                             & \textbf{7}                                                      & 4                                                           & 5               \\
\multicolumn{1}{l|}{Function Coverage}      & \textbf{8}                                                    & 7                                                               & 6                                                           & 7               \\
\multicolumn{1}{l|}{Line Coverage}          & 4                                                             & \textbf{6}                                                      & 4                                                           & 5               \\
\multicolumn{1}{l|}{Region Coverage}        & 4                                                             & \textbf{6}                                                      & 4                                                           & \textbf{6}      
\\ \hline \hline
\end{tabular}
\end{table}
Additionally, \emph{Table \ref{table:cip-branch}} reports the percentage gain (i.e., CIP) of the \emph{best-performing} LLM-guided fuzzer over the AFL++ baseline for each prompt shot, based on branch coverage. 
Among the six benchmarks evaluated for branch coverage with 3-shot prompts, \emph{four benchmarks} show positive CIPs in one-hour runs, indicating that LLM-guided fuzzers generally outperform the AFL++ baseline in short runs, though this trend does not always hold for longer runs.

\begin{table*}[]
\centering
\caption{Branch coverage for multiple LLMs comparison experiments.\label{long}}
\label{table:cip-branch}
\begin{tabular}{llllrllr}
\hline \hline
\multicolumn{8}{c}{\textbf{Branch Coverage: Multiple LLMs Comparison Experiment}} \\
\hline \hline
\multicolumn{1}{l|}{\multirow{2}{*}{\textbf{Shot}}} & \multicolumn{1}{l|}{\multirow{2}{*}{\textbf{Benchmarks}}} & \multicolumn{3}{l|}{\textbf{One-Hour Runtime}}                                                                                                                                                                 & \multicolumn{3}{l}{\textbf{Four-Hour Runtime}}                                                                                                                                                                \\ \cline{3-8} 
\multicolumn{1}{l|}{}                               & \multicolumn{1}{l|}{}                                     & \textbf{\begin{tabular}[c]{@{}l@{}}LLM\\ (\%)\end{tabular}} & \textbf{\begin{tabular}[c]{@{}l@{}}AFL++\\ (\%)\end{tabular}} & \multicolumn{1}{l|}{\textbf{\begin{tabular}[c]{@{}l@{}}CIP\\ (\%)\end{tabular}}} & \textbf{\begin{tabular}[c]{@{}l@{}}LLM\\ (\%)\end{tabular}} & \textbf{\begin{tabular}[c]{@{}l@{}}AFL++\\ (\%)\end{tabular}} & \multicolumn{1}{l}{\textbf{\begin{tabular}[c]{@{}l@{}}CIP\\ (\%)\end{tabular}}} \\ \hline
\multicolumn{1}{l|}{\multirow{6}{*}{0}}             & \multicolumn{1}{l|}{freetype2\_ftfuzzer}                  & 37.66                                                       & 36.94                                                         & \multicolumn{1}{r|}{0.72}                                                        & 39.47                                                       & 38.72                                                         & 0.75                                                                            \\
\multicolumn{1}{l|}{}                               & \multicolumn{1}{l|}{libpng\_libpng\_read\_fuzzer}         & 32.72                                                       & 33.17                                                         & \multicolumn{1}{r|}{-0.45}                                                       & 33.32                                                       & 33.57                                                         & -0.25                                                                           \\
\multicolumn{1}{l|}{}                               & \multicolumn{1}{l|}{curl\_curl\_fuzzer\_http}             & 12.78                                                       & 13.07                                                         & \multicolumn{1}{r|}{-0.29}                                                       & 12.76                                                       & 13.35                                                         & -0.59                                                                           \\
\multicolumn{1}{l|}{}                               & \multicolumn{1}{l|}{libpcap\_fuzz\_both}                  & 36.48                                                       & 37.83                                                         & \multicolumn{1}{r|}{-1.35}                                                       & 35.66                                                       & 40.24                                                         & -4.58                                                                           \\
\multicolumn{1}{l|}{}                               & \multicolumn{1}{l|}{openh264\_decoder\_fuzzer}            & 75.05                                                       & 74.52                                                         & \multicolumn{1}{r|}{0.53}                                                        & 76.48                                                       & 76.10                                                         & 0.38                                                                            \\
\multicolumn{1}{l|}{}                               & \multicolumn{1}{l|}{openthread\_ot-ip6-send-fuzzer}       & 11.23                                                       & 11.41                                                         & \multicolumn{1}{r|}{-0.18}                                                       & 11.28                                                       & 11.50                                                         & -0.22                                                                           \\ \hline
\multicolumn{1}{l|}{\multirow{6}{*}{1}}             & \multicolumn{1}{l|}{freetype2\_ftfuzzer}                  & 37.27                                                       & 36.94                                                         & \multicolumn{1}{r|}{0.33}                                                        & 40.01                                                       & 38.72                                                         & 1.29                                                                            \\
\multicolumn{1}{l|}{}                               & \multicolumn{1}{l|}{libpng\_libpng\_read\_fuzzer}         & 33.01                                                       & 33.17                                                         & \multicolumn{1}{r|}{-0.16}                                                       & 32.82                                                       & 33.57                                                         & -0.75                                                                           \\
\multicolumn{1}{l|}{}                               & \multicolumn{1}{l|}{curl\_curl\_fuzzer\_http}             & 12.79                                                       & 13.07                                                         & \multicolumn{1}{r|}{-0.28}                                                       & 12.67                                                       & 13.35                                                         & -0.68                                                                           \\
\multicolumn{1}{l|}{}                               & \multicolumn{1}{l|}{libpcap\_fuzz\_both}                  & 39.08                                                       & 37.83                                                         & \multicolumn{1}{r|}{1.25}                                                        & 38.22                                                       & 40.24                                                         & -2.02                                                                           \\
\multicolumn{1}{l|}{}                               & \multicolumn{1}{l|}{openh264\_decoder\_fuzzer}            & 75.31                                                       & 74.52                                                         & \multicolumn{1}{r|}{0.79}                                                        & 76.41                                                       & 76.10                                                         & 0.31                                                                            \\
\multicolumn{1}{l|}{}                               & \multicolumn{1}{l|}{openthread\_ot-ip6-send-fuzzer}       & 11.2                                                        & 11.41                                                         & \multicolumn{1}{r|}{-0.21}                                                       & 11.34                                                       & 11.50                                                         & -0.16                                                                           \\ \hline
\multicolumn{1}{l|}{\multirow{6}{*}{3}}             & \multicolumn{1}{l|}{freetype2\_ftfuzzer}                  & 37.34                                                       & 36.94                                                         & \multicolumn{1}{r|}{0.40}                                                        & 40.26                                                       & 38.72                                                         & 1.54                                                                            \\
\multicolumn{1}{l|}{}                               & \multicolumn{1}{l|}{libpng\_libpng\_read\_fuzzer}         & 32.95                                                       & 33.17                                                         & \multicolumn{1}{r|}{-0.22}                                                       & 32.80                                                       & 33.57                                                         & -0.77                                                                           \\
\multicolumn{1}{l|}{}                               & \multicolumn{1}{l|}{curl\_curl\_fuzzer\_http}             & 12.79                                                       & 13.07                                                         & \multicolumn{1}{r|}{-0.28}                                                       & 12.80                                                       & 13.35                                                         & -0.55                                                                           \\
\multicolumn{1}{l|}{}                               & \multicolumn{1}{l|}{libpcap\_fuzz\_both}                  & 37.91                                                       & 37.83                                                         & \multicolumn{1}{r|}{0.08}                                                        & 34.99                                                       & 40.24                                                         & -5.25                                                                           \\
\multicolumn{1}{l|}{}                               & \multicolumn{1}{l|}{openh264\_decoder\_fuzzer}            & 75.31                                                       & 74.52                                                         & \multicolumn{1}{r|}{0.79}                                                        & 76.34                                                       & 76.10                                                         & 0.24                                                                            \\
\multicolumn{1}{l|}{}                               & \multicolumn{1}{l|}{openthread\_ot-ip6-send-fuzzer}       & 11.43                                                       & 11.41                                                         & \multicolumn{1}{r|}{0.02}                                                        & 11.34                                                       & 11.50                                                         & -0.16                                              \\ \hline \hline
\end{tabular}
\end{table*}

\begin{figure}[t]
    \centering
    \includegraphics[width=0.45\textwidth]{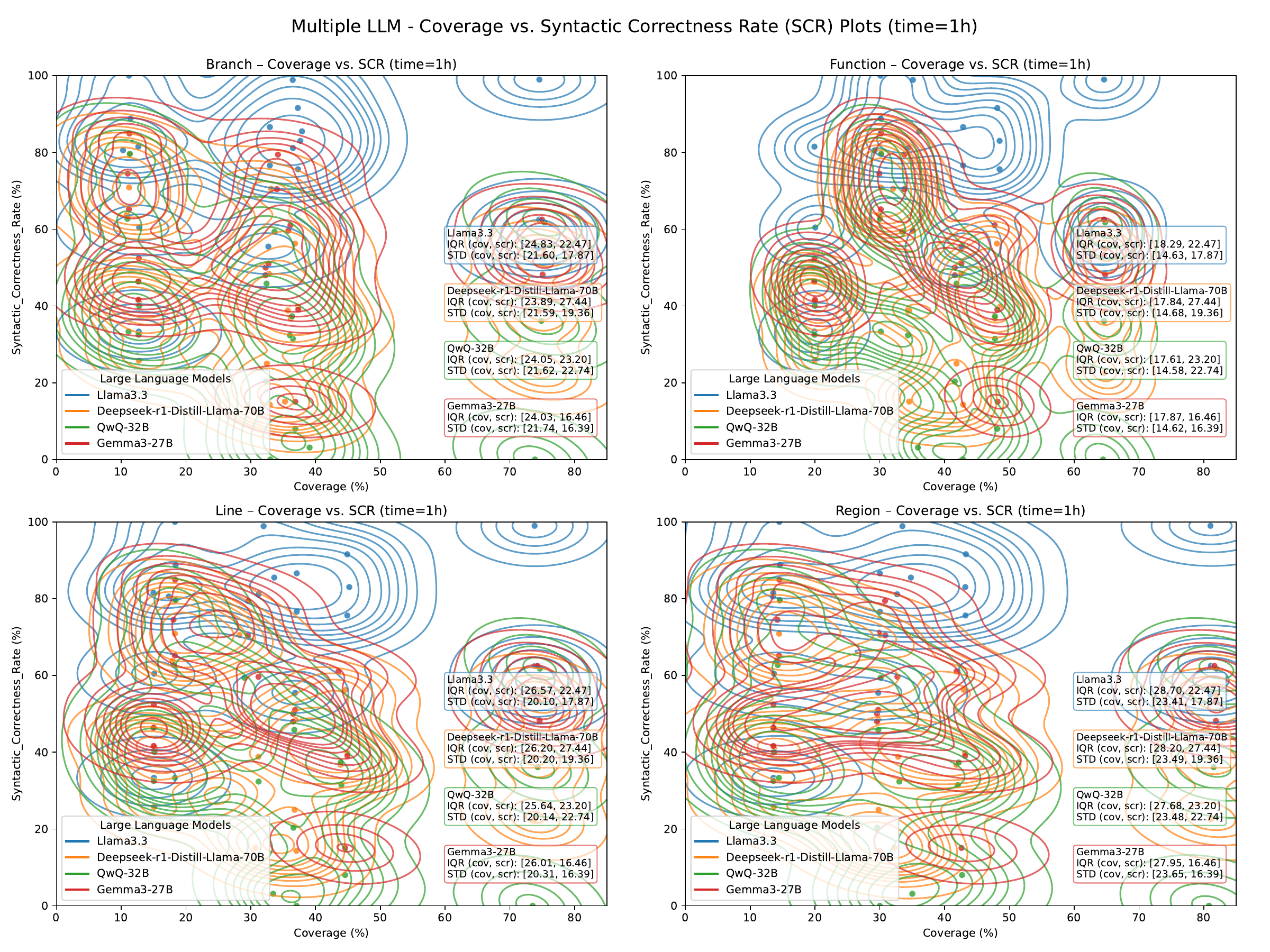}
    \caption{Multiple LLMs Coverage vs SCR for one-hour runtime}
    \label{Fig:llm-1-cs-kde}
\end{figure}

\begin{figure}[t]
    \centering
    \includegraphics[width=0.45\textwidth]{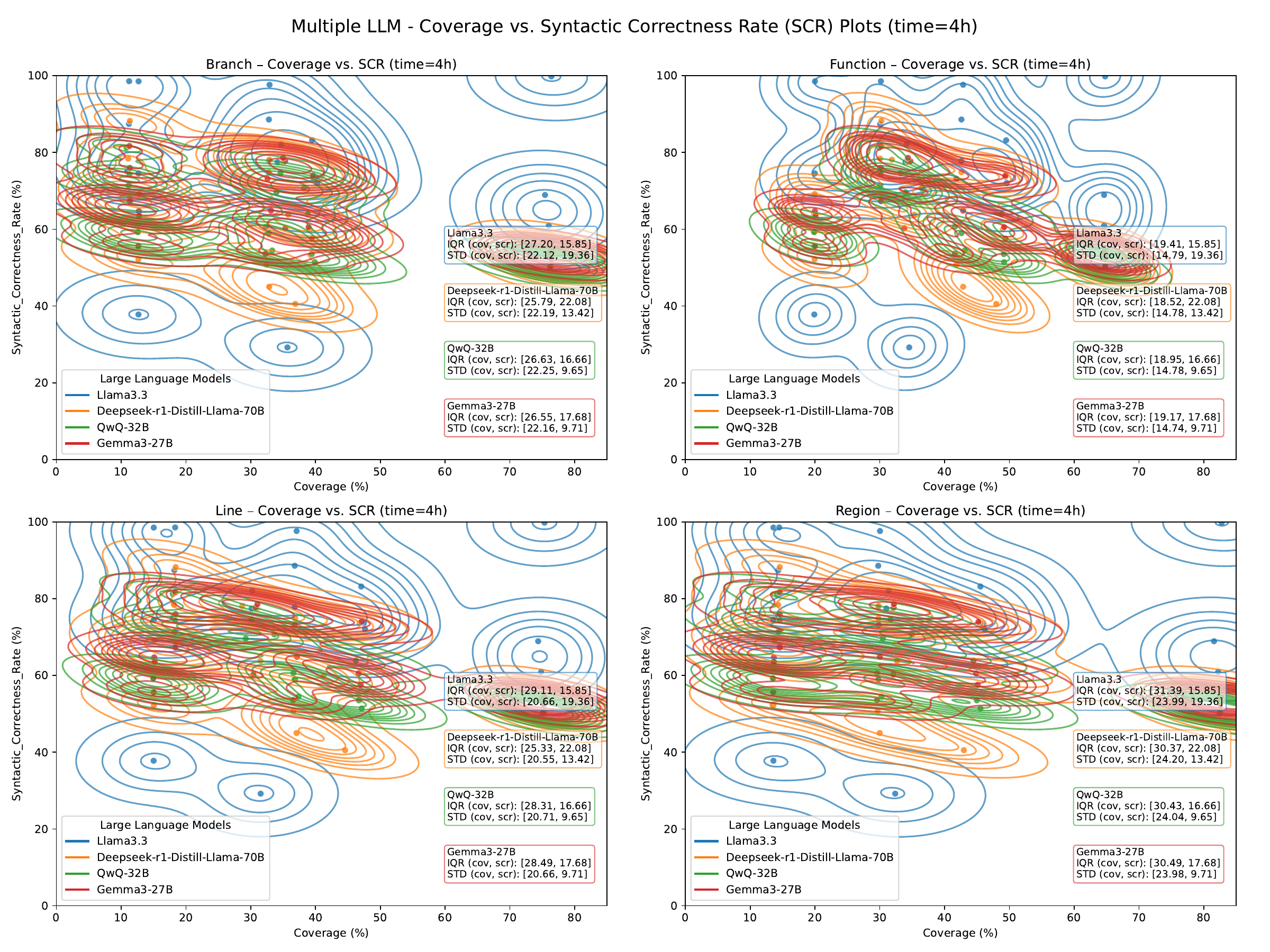}
    \caption{Multiple LLMs Coverage vs SCR for four-hour runtime}
    \label{Fig:llm-4-cs-kde}
\end{figure}

\begin{figure}[t]
    \centering
    \includegraphics[width=0.45\textwidth]{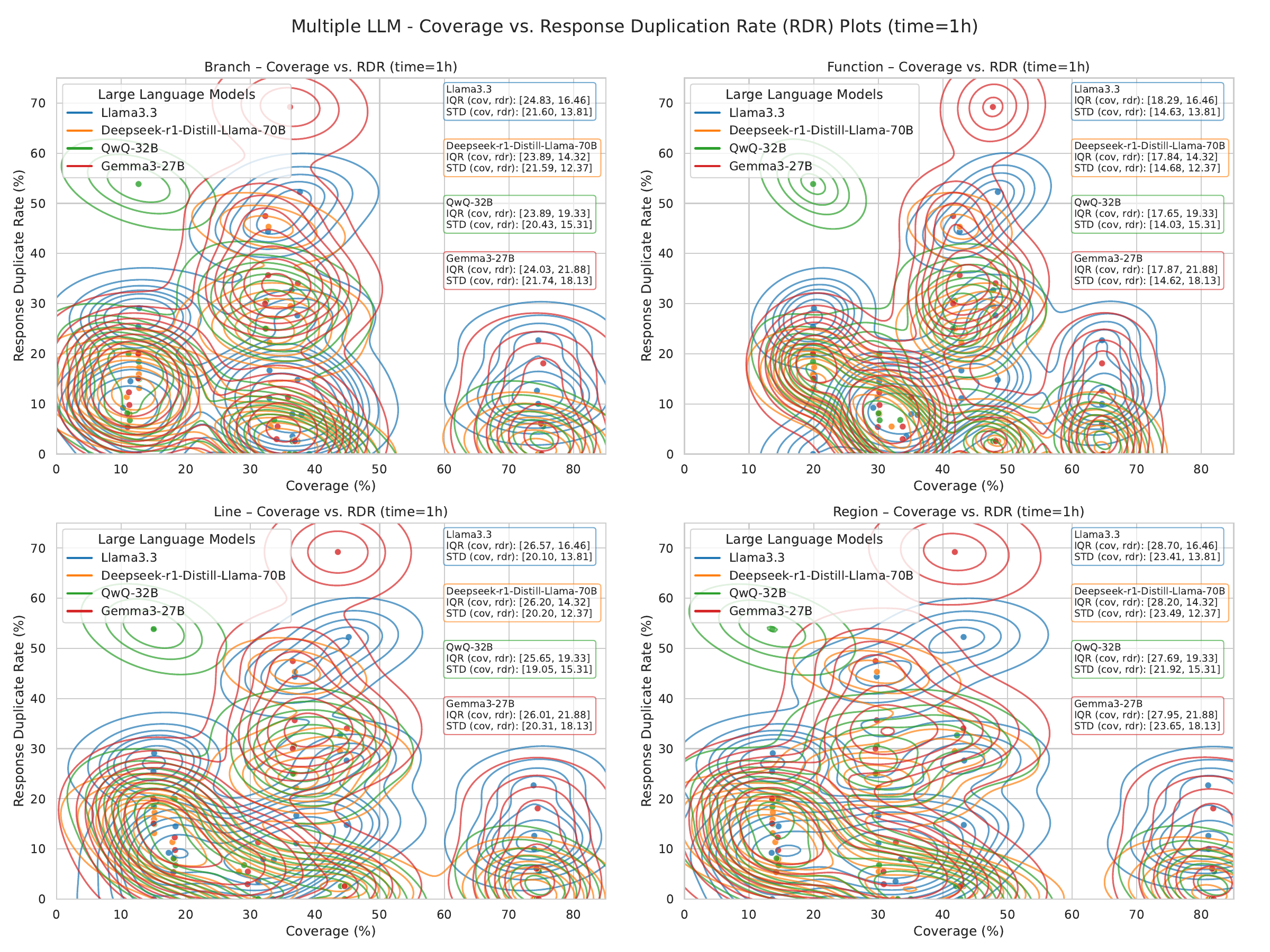}
    \caption{Multiple LLMs Coverage vs RDR for one-hour runtime}
    \label{Fig:llm-1-cr-kde}
\end{figure}

\begin{figure}[t]
    \centering
    \includegraphics[width=0.45\textwidth]{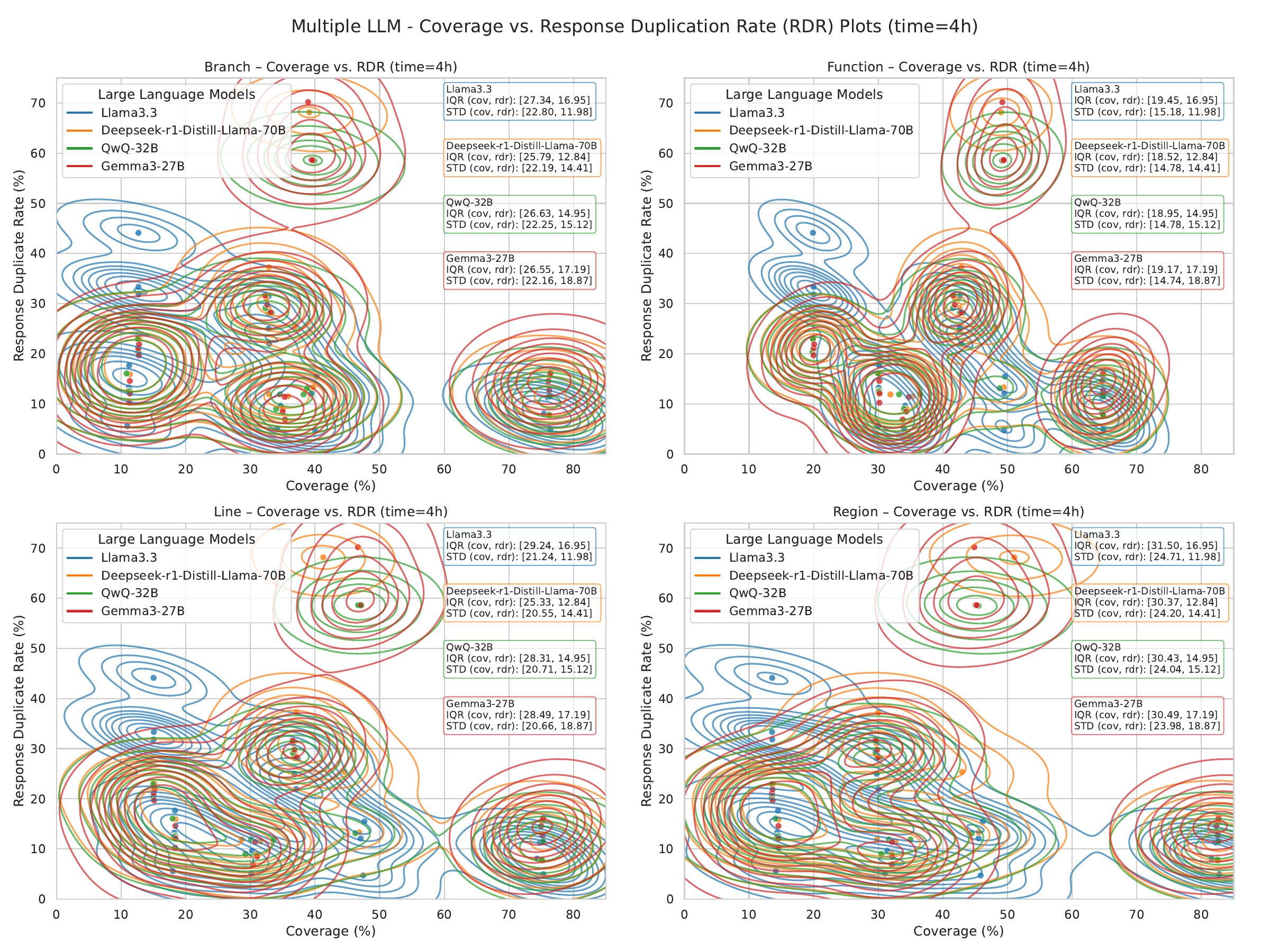}
    \caption{Multiple LLMs Coverage vs RDR for four-hour runtime}
    \label{Fig:llm-4-cr-kde}
\end{figure}

To determine which LLM performs best overall and answer \textbf{R4}, we next analyze model response quality in terms of syntactic correctness (SCR) and response diversity (RDR). We first evaluated the \emph{SCR} and \emph{RDR} metrics for each LLM across all three prompt shots, then plotted code coverage for all selected LLMs against these metrics.
\emph{Figures \ref{Fig:llm-1-cs-kde} and \ref{Fig:llm-4-cs-kde}} visualize the relationship between coverage and SCR across LLMs and prompt shots;
and \emph{Figures \ref{Fig:llm-1-cr-kde} and \ref{Fig:llm-4-cr-kde}} present the correlation between coverage and RDR.

The SCR plots reveal that all models achieve higher syntactic correctness alongside improved code coverage.
\emph{Llama3.3} consistently maintains \emph{the strongest SCR-coverage correlation} in one-hour runs, with its KDE plots concentrated in high SCR and moderate-to-high coverage regions. However, its SCR declines when the fuzzing period is extended to four hours, suggesting that formatting consistency may degrade over time and could benefit from runtime constraints. \emph{Deepseek-r1-Distill-Llama-70B} shows higher SCR variability with a slightly narrower coverage spread than Llama3.3, indicating \emph{higher sensitivity to response fluctuations}. Despite unstable syntactic correctness in longer runs, the model still maintains \emph{consistent coverage}.
\emph{Gemma3-27B} shows low SCR and scattered KDE distributions across prompt shots and coverage types in one-hour runs, indicating \emph{weaker formatting consistency} and less promising early-stage coverage than Deepseek. 
\emph{QwQ-32B} offers relatively balanced coverage with moderate variability, but its low SCR values in four-hour runs indicate limited formatting reliability.

In the RDR plots, \emph{Deepseek-r1-Distill-Llama-70B} consistently achieves low RDR—especially in long-term fuzzing. It shows a relatively high SCR variability but centralized low-RDR densities, indicating frequent production of diverse and well-formatted inputs that sustain stable code coverage performance.
In contrast, \emph{Gemma3-27B} demonstrates scattered RDR values, suggesting frequent generation of repetitive responses. \emph{Llama3.3} shows low RDR but with greater coverage variability compared to Deepseek. \emph{QwQ-32B} remains mid-range in both diversity and coverage.

\subsection{Discussion}
After evaluating experimental results using four key metrics—detailed code coverage, CIP, SCR, and RDR—we analyzed trends to explore how LLM syntactic correctness and response diversity influence fuzzing performance. Together, these metrics establish a comprehensive basis for addressing our research questions. The following sections highlight key findings drawn from the tables and plots in the results. 

\subsubsection{Observation 1: Findings Related to R2
}
The Llama3.3 prompt-engineering experiments reveal that \emph{increasing prompt shots does not consistently improve code coverage}. While some benchmarks benefit from additional in-context examples, others show negligible or even negative effects. This suggests that prompt complexity does not linearly correlate with fuzzing effectiveness. In cases where coverage improves, extended examples in the prompt may enhance syntactic correctness of LLM responses. Conversely, coverage declines in some benchmarks may result from increased predictability and reduced exploratory variation in the LLM's outputs. These trade-offs highlight the need for a \emph{balance between prompt structure and model variability}, which is further explored in subsequent discussions.

\subsubsection{Observation 2: Findings Related to R3
}
CIP results demonstrate that \emph{no single LLM consistently outperforms the AFL++ baseline} across all benchmarks and prompt shots. 
While some LLMs exceed AFL++'s performance on specific benchmarks, there are also many cases where the LLMs fall short, suggesting that current state-of-the-art LLMs are not universally superior to traditional fuzzers. 
Among the models, Llama3.3 shows \emph{the strongest direct correlation} between syntactic correctness and coverage, achieving the \emph{best performance} in early fuzzing. While Deepseek-r1-Distill-Llama-70B is \emph{highly sensitive} to response diversity, it consistently generates mutated inputs in stable formats and performs best in long-term fuzzing experiments. Gemma3-27B and QwQ-32B both perform moderately in longer runs—Gemma3 offering more stability and QwQ showing weaker formatting consistency—making them suitable as mid-tier baselines for testing predictable results.
Overall, \emph{Llama3.3} and \emph{Deepseek-r1-Distill-Llama-70B} are strong candidates for further research: Llama3.3 for early-phase fuzzing with low error variance, and Deepseek for long-term vulnerability explorations.

\subsubsection{Observation 3: Impact of Syntactic Correctness on Code Coverage
}
Both the multi-LLM comparison and the Llama3.3 prompt-engineering experiments examined the relationship between code coverage and LLM syntactic correctness (i.e., SCR). The results show that \emph{higher code coverage generally aligns with higher SCR values}, while increasing prompt shots stabilize SCR values and may improve coverage. This suggests that \emph{better syntactic correctness enables more usable mutated inputs and improved path exploration}.
However, this effect follows a \emph{threshold pattern}: Beyond a certain point, further improvements in SCR do not result in additional code coverage gains, forming an "L" shape in the SCR-coverage relationship plots. These findings indicate that while syntactic correctness is essential, it alone is \emph{insufficient to continuously improve fuzzing efficiency}, highlighting the need to consider additional factors to enhance fuzzing performance.

\subsubsection{Observation 4: Impact of Response Diversity on Code Coverage
}
LLM response diversity, measured by RDR, is another key factor affecting fuzzing effectiveness. The RDR-coverage relationship plots reveal an inverse relationship between RDR and code coverage: Lower RDR generally correlates with higher coverage, indicating that \emph{greater mutation diversity produces more effective test cases}. Increasing prompt shots in one-hour runs tends to raise RDR values, as the LLM may produce more deterministic outputs. In four-hour runs, RDR varies widely across prompt shots while coverage remains relatively stable, showing that additional in-context examples can increase duplication without improving coverage. These observations highlight a potential \emph{trade-off}: Prompt engineering can improve syntactic correctness but reduce response diversity. 
Balancing response syntactic correctness and diversity is therefore essential for optimizing LLM-guided fuzzing performance.

\subsubsection{Observation 5: Additional Impacts in Fuzzing Results}
Log analysis shows that approximately \emph{35\%} of LLM queries timed out using \emph{Ollama}, particularly with longer prompts or more prompt shots. This timeout behavior limits the consistent delivery of LLM-generated mutations in real time, reducing the number of effective mutations during fuzzing. While our system uses AFL++'s default mutation strategies to maintain operation when the LLM fails to respond, this can result in fewer novel paths and diminish the benefits of LLM guidance. 
These findings highlight the importance of LLM responsiveness and stability in our solution.

\section{Conclusion and Future Work}
We proposed a new mutation-based fuzzer that integrates off-the-shelf reasoning LLMs into AFL++ within a microservices architecture, enabling large-scale, reproducible experiments via Google’s Fuzzbench benchmarking platform. Our framework addresses \textbf{Challenge 1-4} in system integration and prompt structuring (\textbf{R1}), providing an \emph{empirical study} for evaluating state-of-the-art open-source reasoning LLMs in mutation-based fuzzing. 
Our study systematically examined effects of prompt engineering—using zero-, one-, and three-shot learning—along with LLM inference on mutation quality and fuzzing efficiency.  
Key findings include:
\squishlist
        \item \textbf{Prompt Shots (R2):} Increasing prompt shots \emph{does not} linearly improve fuzzing. Higher-shot prompts can improve syntactic correctness but may also make LLM outputs overly deterministic, reducing mutation diversity. Thus, balancing prompt design with model behavior is key for effective LLM-guided fuzzing.
        \item \textbf{Reasoning LLM Performance (R3):} No \emph{single reasoning LLMs} consistently outperform traditional fuzzers in mutation generation without fine-tuning or additional training. However, \emph{Llama3.3} and \emph{Deepseek-r1-Distill-Llama-70B} show strong potential to enhance mutation effectiveness. 
        \item \textbf{Best LLM Selection (R4):} \emph{Deepseek-r1-Distill-Llama-70B} achieves the best balance between output diversity and syntactic correctness, making it the top candidate for improving mutation quality and code coverage in long-term fuzzing within our architecture.
        \item \textbf{Syntactic Correctness vs. Response Diversity:} Effective LLM-guided fuzzing requires balancing syntactic correctness and response diversity. While stable SCR ensures valid mutations, higher SCR levels do not increase coverage beyond a threshold. Improved response diversity enhances fuzzing efficiency, emphasizing that well-formatted and diverse LLM outputs are essential for LLM-guided fuzzing. 
        \item \textbf{LLM Response Latency and Timeouts:} Lengthy or complex prompts can cause LLM response latency or timeouts, reducing effective mutations and limiting the benefits of LLM integration. Scalable fuzzing requires optimizing LLM responsiveness and deployment infrastructure.
\squishlistend

Despite promising results, our approach has five limitations: (1) Mutations sometimes deviate from required formats, (2) LLMs lack native binary support, (3) large inputs can exceed token limits or trigger timeouts, (4) the experimental scale was constrained by time and resource, and (5) Fuzzbench retains detailed reports for only one trial—limiting multi-trial analysis. Future work will explore fine-tuning LLMs—guided by automated mutation feedback and implemented through reinforcement learning or Direct Preference Optimization (DPO)—to improve both syntactic correctness and mutation diversity, thereby generating more effective inputs for LLM-guided fuzzing. Other directions include optimizing the Fuzzbench pipeline and LLM serving infrastructure, exploring input chunking strategies, and extending tests to longer fuzzing durations. These efforts aim to enhance the scalability, robustness, and effectiveness of LLM-guided mutation-based fuzzing.

\section{Related Work}
Traditional grey-box fuzzers rely on random or heuristic input mutations to discover vulnerabilities in target programs through brute-force testing \cite{sutton2007what}. While AFL \cite{10.1145/3580596} and AFL++ \cite{10.5555/3488877.3488887} improved efficiency with feedback-driven evolutionary algorithms, their mutation strategies are still shallow, surface-level edits that limit deeper vulnerability discovery. LibFuzzer \cite{libfuzzerDocs} and SelectFuzz \cite{10179296} optimize instrumentation to monitor code coverage and leverage the feedback to guide input mutations, but they still use static, heuristic-based mutation strategies which struggle with inputs requiring complex syntactic or semantic constraints and often generate invalid or ineffective inputs \cite{limit1, limit2}. These challenges hinder deeper bug discovery—especially in modern software systems with evolving logic and formats caused by frequent implementation changes \cite{10.1145/3597503.3639121}.

To overcome these limitations, researchers have explored intelligent input prioritization and mutation techniques \cite{neuzz, 10580893, 9199813}, leading to integrating machine learning and, more recently, LLMs into fuzzing workflows. ML-enhanced fuzzers such as V-Fuzz \cite{9199813} and CTFuzz \cite{10580893} use neural networks or reinforcement learning to prioritize inputs and guide mutation strategies, improving efficiency over random mutations. The rise of LLMs further enables semantic understanding of input formats and mutation guidance, with two main approaches: (1) \emph{Fine-tuning} \cite{finetune} the model, which adapts LLMs through supervised training on domain-specific input; and (2) \emph{prompt engineering} \cite{KNOTH2024100225}, which crafts structured prompts at inference time without retraining. 
However, fine-tuning requires labeled data and restricts mutations to existing heuristics at high computational costs. For instance, LLAMAFUZZ \cite{zhang2024llamafuzz} outperforms AFL++ on select benchmarks but struggles to generalize across new programs, formats, or domains. Conversely, prompt engineering is lightweight and flexible, but current fuzzers—such as Fuzz4All \cite{10.1145/3597503.3639121}, PromptFuzz  \cite{10.1145/3658644.3670396}, and CHATAFL \cite{meng2024llmfuzz}—treat LLMs as black-box, input-output generators, offering little insight into how mutations are derived. 

Reasoning-enable LLMs offer a promising new direction by making the mutation process more transparent. Unlike prior prompt-only methods, reasoning LLMs such as Llama3 \cite{2024arXiv240721783G}, Deepseek-r1 \cite{bau2024gradient}, and Gemma3 \cite{gemma3} generate a logical progression of reasoning, or "chain-of-thought", that explains how the final output is derived. This capability not only supports better prompt design but also reduces redundant or invalid mutations and enables deeper path exploration. Despite this potential, reasoning-driven fuzzing remains underexplored: no prior work has systematically benchmarked reasoning LLMs across diverse targets or investigated how prompt-shot strategies influence mutation diversity and syntactic correctness. Addressing this gap, our research integrates reasoning LLMs with AFL++ in the FuzzBench framework, providing the first empirical analysis of prompt-shot learning, response duplication, and syntactic correctness in reasoning LLM-guided fuzzing.

\appendices

\ifCLASSOPTIONcaptionsoff
  \newpage
\fi

\bibliographystyle{IEEEtran}
\bibliography{ms}

\end{document}